\documentclass[prb,showpacs,floatfix,twocolumn,byrevtex]{revtex4}
\usepackage{dcolumn}
\usepackage{graphicx}
\usepackage{float}
\usepackage{longtable}
\begin{document}

\bibliographystyle{apsrev}

\preprint{Draft version 2.0, not for distribution}
%
% The title and the list of authors
%
\title{Scaling of the superfluid density in high-temperature superconductors}

\author{C. C. Homes}
\email{homes@bnl.gov}
\author{S. V. Dordevic}
\author{T. Valla}
\author{M. Strongin}
\affiliation{Department of Physics, Brookhaven National Laboratory, Upton, New
York 11973}%
%
%\author{M. Greven}
%\affiliation{Stanford Synchrotron Radiation Laboratory, Stanford, CA 94309}
%\affiliation{Department of Applied Physics, Stanford University, Stanford,
%California 94305}
%
\date{\today}

%
% The abstract goes here
%
\begin{abstract}
A scaling relation $\rho_s \simeq 35\,\sigma_{dc}\,T_c$ has been observed in
the copper-oxide superconductors, where $\rho_s$ is the strength of the
superconducting condensate, $T_c$ is the critical temperature, and
$\sigma_{dc}$ is the normal-state dc conductivity  close to $T_c$.  This
scaling relation is examined within the context of a clean and dirty-limit BCS
superconductor.  These limits are well established for an isotropic BCS gap
$2\Delta$ and a normal-state scattering rate $1/\tau$; in the clean limit
$1/\tau \ll 2\Delta$, and in the dirty limit $1/\tau > 2\Delta$. The dirty
limit may also be defined operationally as the regime where $\rho_s$ varies
with $1/\tau$.  It is shown that the scaling relation $\rho_s \propto
\sigma_{dc}\,T_c$ is the hallmark of a BCS system in the dirty-limit. While the
gap in the copper-oxide superconductors is considered to be {\it d}-wave with
nodes and a gap maximum $\Delta_0$, if $1/\tau > 2\Delta_0$ then the
dirty-limit case is preserved. The scaling relation implies that the
copper-oxide superconductors are likely to be in the dirty limit, and that as a
result the energy scale associated with the formation of the condensate is
scaling linearly with $T_c$.  The {\it a-b} planes and the {\it c} axis also
follow the same scaling relation.  It is observed that the scaling behavior for
the dirty limit and the Josephson effect (assuming a BCS formalism) are
essentially identical, suggesting that in some regime these two effects may be
viewed as equivalent.  This raises the possibility that electronic
inhomogeneities in the copper-oxygen planes may play an important role in the
nature of the superconductivity in the copper-oxide materials.
\end{abstract}
%
%  PACS numbers
%  63.20.-e  Phonons in crystal lattices
%  72.15.Lh  Relaxation times and mean-free paths
%  74.25.Gz  Optical properties
%  74.25.-q  Normal and superconducting states
%  74.72.-h  High-Tc compounds
%  74.72.Bk  Y-based cuprates
%  77.22.Ch  Permittivity (dielectric function)
%  78.30.-j  Infrared and Raman spectra
%
\maketitle

%
% The main body of the text
%
% Introduction
%
\section{Introduction}
\begin{figure}[t]%
%
% eprint
%
%\vspace*{-0.5cm}%
\centerline{\includegraphics[width=3.4in]{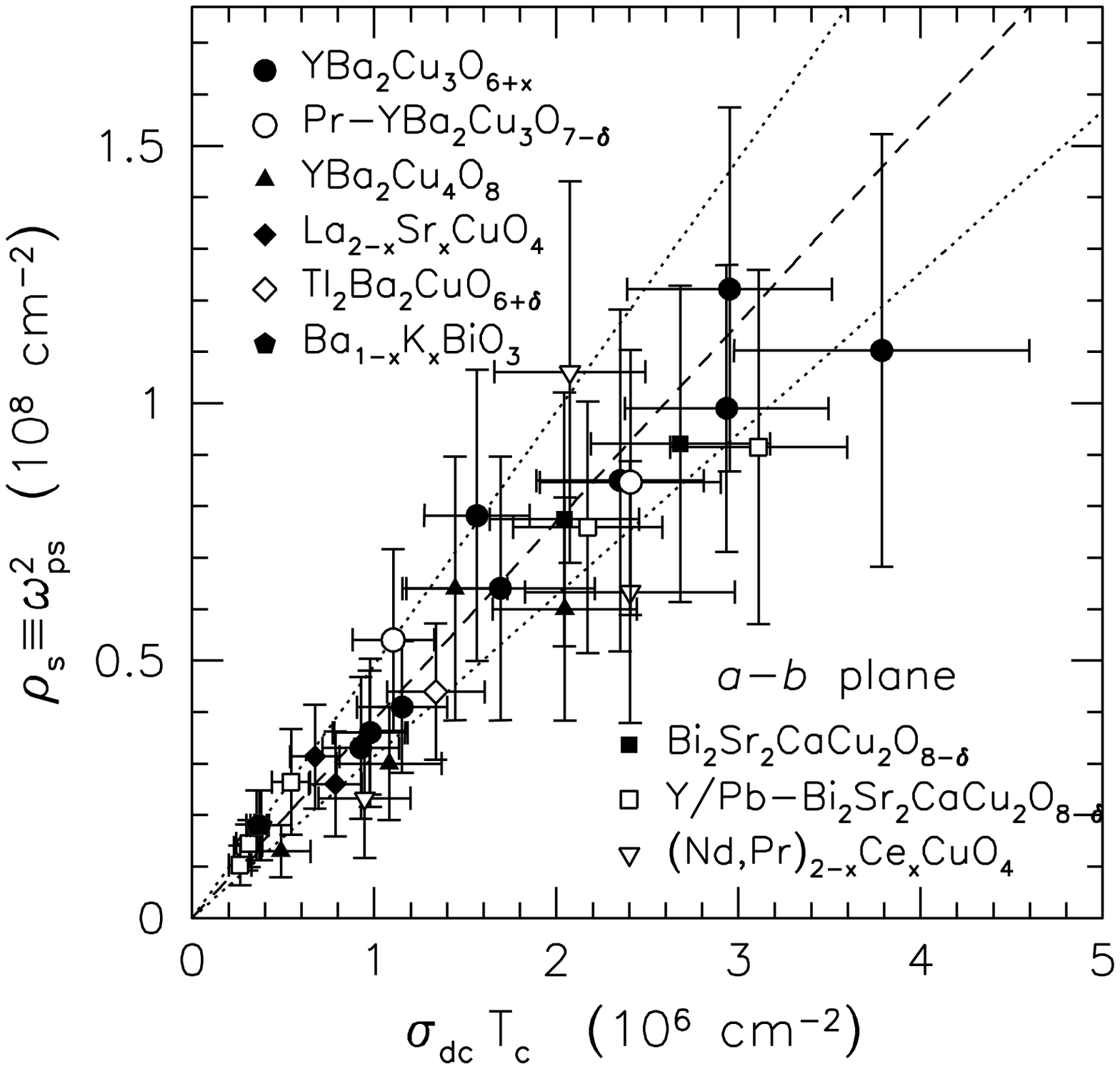}}%
\vspace*{-0.3cm}%
%
% manuscript
%
%\centerline{\includegraphics[width=6.0in]{figure1.eps}}%
%\centerline{\includegraphics[width=6.7in]{color1.eps}}%
%\vspace*{-0.3cm}%
%
\caption{The superfluid density $\rho_s$ vs $\sigma_{dc}\,T_c$ for the {\it a-b}
planes of the hole-doped copper-oxide superconductors for pure and Pr-doped
YBa$_2$Cu$_3$O$_{6+x}$
(Refs.~\onlinecite{basov94,basov95a,homes99,pimenov99,liu99}); YBa$_2$Cu$_4$O$_8$
(Ref.~\onlinecite{basov95a}); pure and Y/Pb-doped Bi$_2$Sr$_2$CaCu$_2$O$_{8+\delta}$
(Refs.~\onlinecite{liu99,wang99,tu02}); underdoped La$_{2-x}$Sr$_x$CuO$_4$
(Ref.~\onlinecite{startseva99}); Tl$_2$Ba$_2$CuO$_{6+\delta}$
(Ref.~\onlinecite{puchkov95}); electron-doped (Nd,Pr)$_{2-x}$Ce$_x$CuO$_4$
(Refs.~\onlinecite{homes97,singley01,zimmers04}) and the bismate material
Bi$_{1-x}$K$_x$BiO$_3$ (Ref.~\onlinecite{puchkov96b}).  Within error, all the points
may be described by a single (dashed) line, $\rho_s \simeq 35\,\sigma_{dc}\,T_c$;
the upper and lower dotted lines, $\rho_s\simeq 44\,\sigma_{dc}\,T_c$ and
$28\,\sigma_{dc}\,T_c$ respectively, represent approximately the spread of the data.}%
\vspace*{-0.2cm}%
\label{fig:linear}
\end{figure}

Scaling laws express a systematic and universal simplicity among complex
systems in nature. For example, such laws are of enormous significance in
biology, where the scaling relation between body mass and metabolic rate spans
21 orders of magnitude.\cite{whittaker99,brown99}  Scaling relations are
equally important in the physical sciences. Since the discovery of
superconductivity at elevated temperatures in copper-oxide
materials\cite{bednorz86} there has been considerable effort to find trends and
correlations between the physical quantities, as a clue to the origin of the
superconductivity.\cite{schneider02}  One of the earliest patterns that emerged
was the linear scaling of the superfluid density $\rho_s \propto 1/\lambda^2$
(where $\lambda$ is the superconducting penetration depth) in the copper-oxygen
planes of the hole-doped materials with the superconducting transition
temperature $T_c$. This is referred to as the Uemura
relation,\cite{uemura89,uemura91} and it works reasonably well for the
underdoped materials.  However, it does not describe very
underdoped,\cite{zuev04} optimally doped (i.e., $T_c$ is a maximum),
overdoped,\cite{niedermayer93,tallon03} or electron-doped
materials.\cite{homes97,homes98a} A similar attempt to scale $\rho_s$ with the
dc conductivity $\sigma_{dc}$ was only partially successful.\cite{pimenov99}
We have recently demonstrated that the scaling relation $\rho_s \propto
\sigma_{dc}\,T_c$ may be applied to a large number of high-temperature
superconductors, regardless of doping level or type, nature of disorder,
crystal structure, or direction (parallel or perpendicular to the copper-oxygen
planes).\cite{homes04b}  The optical values of $\rho_s (T\ll T_c)$ and
$\sigma_{dc} (T\gtrsim T_c)$ within the {\it a-b} planes have been determined
for a large number of copper-oxide superconductors, as well as the
bismuth-oxide material Ba$_{1-x}$K$_x$BiO$_3$; the results are shown as a
linear plot in Fig.~\ref{fig:linear}.  In this representation, the underdoped
points near the origin tend to lie rather close together, and the {\it c}-axis
points which are not shown would be visible only as a single point slightly
below the underdoped data. While there is some scatter in the data, a linear
trend is clearly visible, although it has been suggested that there is some
deviation from this behavior in the extremely-overdoped
materials.\cite{tallon04}
When plotted as a log-log plot in Fig.~\ref{fig:abplane}, the linear trend is
more apparent, and indicates that within the error the points may be described
by the scaling relation $\rho_s \simeq 35\,\sigma_{dc}\,T_c$ (in this instance
both sides of the equation possess the same units, so that the constant is
dimensionless). In addition, the elemental BCS superconductors Nb and Pb
(without any special regards to preparation) are also observed to follow this
scaling relation reasonably well.

%
% Conductivity and superfluid density
%
\section{Experiment}
The values for $\sigma_{dc}$ and $\rho_s$ shown in Table~I have been obtained
almost exclusively from reflectance measurements from which the complex optical
properties have been determined through a Kramers-Kronig
analysis.\cite{homes93}  The dc conductivity has been extrapolated from the
real part of the optical conductivity $\sigma_{dc} = \sigma_1(\omega
\rightarrow 0)$ at $T\gtrsim T_c$. For $T\ll T_c$, the response of the
dielectric function to the formation of a condensate is expressed purely by the
real part of the dielectric function $\epsilon_1(\omega) = \epsilon_\infty
-\omega_{ps}^2 / \omega^2$, which allows the strength of the condensate to be
calculated from $\omega_{ps}^2 = -\omega^2 \epsilon_1(\omega)$ in the $\omega
\rightarrow 0$ limit. Here, $\omega_{ps}^2=4\pi n_s e^2/m^\ast$ is the square
of the superconducting plasma frequency and $\rho_s \equiv \omega_{ps}^2$, and
$\epsilon_\infty$ is the high-frequency contribution to the real part of the
dielectric function. The strength of the condensate may also be estimated by
tracking the changes in the spectral weight above and below $T_c$, where the
spectral weight is defined as\cite{units}
  $N(\omega,T) = { ({120}/{\pi}) } \int_{0^+}^{\omega} \sigma_1(\omega^\prime,T)
  \,d\omega^\prime$.
The condensate may be calculated from the shift in the spectral weight $\rho_s
= N_n - N_s$, where $N_n = N(\omega, T\simeq T_c)$, and $N_s = N_s(\omega, T\ll
T_c)$. This is the Ferrell-Glover-Tinkham sum rule which tracks changes in the
optical conductivity $\sigma_1(\omega)$ above and below $T_c$ due to the
formation of a condensate at zero frequency.\cite{ferrell58, tinkham59}  These
two different techniques typically yield nearly identical values for $\rho_s$;
an exception exists in the underdoped materials along the {\it c} axis, where
it has been suggested that there is missing spectral weight.\cite{basov99}

%
% Figure 1
%

%
% Scaling relation, clean and dirty limits...
%
\section{Discussion}
A deeper understanding of the scaling relation as it relates to both the elemental
superconductors and the copper-oxide materials may be obtained from an examination
of the spectral weight above and below $T_c$ in relation to the normal-state
scattering rate and the superconducting energy gap. When Nb is in the dirty limit,
it follows the $\rho_s \propto \sigma_{dc}\,T_c$ relation, but in the clean limit
there is a deviation from this linear behavior. (This result will be explored in
more detail shortly.) The terms ``clean'' and ``dirty'' originate from the
comparison of the isotropic BCS energy gap $2\Delta$ with the normal-state
scattering rate $1/\tau$; the clean limit is taken as $1/\tau \ll 2\Delta$, while
the dirty limit is $1/\tau > 2\Delta$. The clean and dirty limits may also be
expressed as $l\gg \xi_0$ and $l < \xi_0$, respectively, where $l$ is the
quasiparticle mean-free path and $\xi_0$ is the BCS coherence length; because
$l\propto \tau$ and $\xi_0 \propto 1/\Delta$, this is equivalent to the previous
statement.\cite{ferrell67}  The use of these definitions depends on having accurate
values for $1/\tau$ and $\Delta$.
%
% Figure 2: log-log plot
%
\begin{figure}[t]%
%
% eprint
%
%\vspace*{-0.5cm}%
\centerline{\includegraphics[width=3.4in]{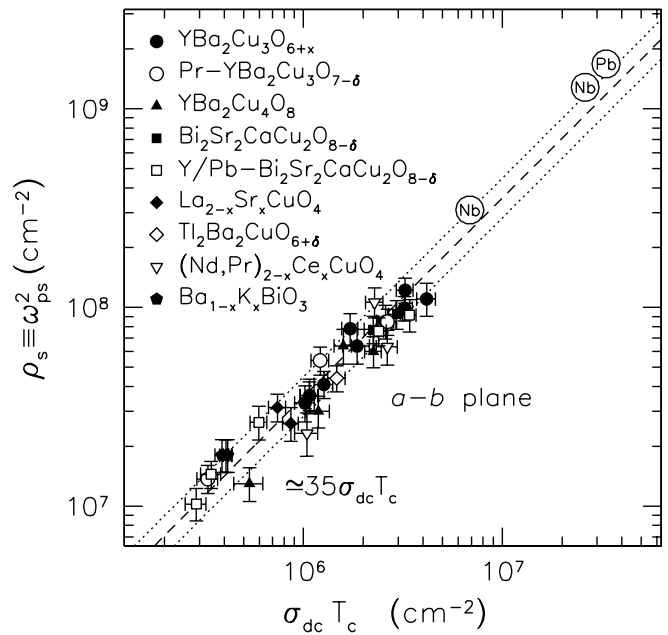}}%
\vspace*{-0.3cm}%
%
% manuscript
%
%\centerline{\includegraphics[width=6.0in]{figure2.eps}}%
%\centerline{\includegraphics[width=6.0in]{figure2.eps}}%
%\vspace*{-0.3cm}%
%
\caption{The log-log plot of the superfluid density $\rho_s$ vs $\sigma_{dc}\,T_c$
for the {\it a-b} planes of the hole-doped copper-oxide superconductors shown in
Fig.~\ref{fig:linear}.  The dashed and dotted lines are shown in this figure are the
sames lines that were shown in Fig.~\ref{fig:linear}. The points for Nb and Pb,
indicated by the atomic symbols, also fall close to the dotted line, $\rho_s \simeq
35\,\sigma_{dc}\,T_c$ (Refs.~\onlinecite{klein94,pronin98}).}%
\vspace*{-0.2cm}%
\label{fig:abplane}
\end{figure}
%
%
%
%
% Table
%
%
% Table 1 - ab plane parameters
%
%\squeezetable{
\begin{table*}
\vspace*{-0.75cm}%
\caption{The critical temperature $T_c$, dc conductivity $\sigma_{dc} \equiv
\sigma_1(\omega \rightarrow 0)$ just above $T\gtrsim T_c$, plasma frequency of the
condensate $\omega_{ps}$ and penetration depth $\lambda_{ab}$ for $T\ll\,T_c$, for
light polarized in the {\it a-b} planes for a variety of single and double-layer
copper-oxygen high-temperature superconductors. Values for Ba$_{1-x}$K$_x$BiO$_3$ as
well as several elemental superconductors have also been included.}
\begin{ruledtabular}
\begin{tabular}{cccccccc}
  Material & Note & (Ref.) &  $T_c$~(K) & $\sigma_{dc}$~($\Omega^{-1}$cm$^{-1}$) &
  $\omega_{ps}$~(cm$^{-1}$) & $\lambda_{ab}$~($\mu$m) \\
%\cline{3-6}
%
% Radiation damage data
%
  & & & & & \\
 YBa$_2$Cu$_3$O$_{6.95}$ & a,b & (\onlinecite{basov94}) & 70   &    $4400\pm 500$  & $5750\pm    600$ & 0.276 \\
 YBa$_2$Cu$_3$O$_{6.95}$ & a,b & (\onlinecite{basov94}) & 80   &    $6500\pm 600$  & $8840\pm    800$ & 0.180 \\
 YBa$_2$Cu$_3$O$_{6.95}$ & a,b & (\onlinecite{basov94}) & 85   &    $9200\pm 900$  & $9220\pm    900$ & 0.172 \\
 YBa$_2$Cu$_3$O$_{6.95}$ & a,b & (\onlinecite{basov94}) & 93.5 & $10\,500\pm 1000$ & $11\,050\pm 800$ & 0.144 \\
 YBa$_2$Cu$_3$O$_{6.60}$ & c &(\onlinecite{basov95a,homes99}) & 59   &    $6500\pm 600$ & $6400\pm 500$ & 0.248 \\
 YBa$_2$Cu$_3$O$_{6.95}$ & c &(\onlinecite{basov95a,homes99}) & 93.2 & $10\,500\pm 900$ & $9950\pm 700$ & 0.159 \\
%
% Liu's data
%
 YBa$_2$Cu$_3$O$_{7-\delta}$    & b & (\onlinecite{liu99}) & 92   &    $8700\pm 900$ & $9200\pm 700$ & 0.172 \\
 Pr-YBa$_2$Cu$_3$O$_{7-\delta}$ & b & (\onlinecite{liu99}) & 40   &    $2500\pm 300$ & $3700\pm 300$ & 0.430 \\
 Pr-YBa$_2$Cu$_3$O$_{7-\delta}$ & b & (\onlinecite{liu99}) & 75   &    $4900\pm 500$ & $7350\pm 600$ & 0.216 \\
 & & & & & \\
%
% Dimitri's Y124 data
%
 YBa$_2$Cu$_4$O$_8$ & c & (\onlinecite{basov95a}) & 80 & $6000\pm 600$ & $8000\pm 800$ & 0.198 \\
 & & & & & \\
%
% Bi2212 data, Y & Pb doped...
%
 Bi$_2$Ca$_2$SrCu$_2$O$_{8+\delta}$      & c & (\onlinecite{tu02}) &  90 & $11\,500 \pm 900$ & $9565\pm 900$ & 0.166 \\
 Bi$_2$Ca$_2$SrCu$_2$O$_{8+\delta}$      & b & (\onlinecite{liu99})  & 91 & $9800\pm 800$ & $9600\pm 800$ & 0.165 \\
 Bi$_2$Ca$_2$SrCu$_2$O$_{8+\delta}$      & b & (\onlinecite{liu99})  & 85 & $8500\pm 800$ & $8710\pm 700$ & 0.182 \\
 Y/Pb-Bi$_2$Ca$_2$SrCu$_2$O$_{8+\delta}$ & b & (\onlinecite{liu99})  & 35 & $2500\pm 300$ & $3200\pm 300$ & 0.497 \\
 Y-Bi$_2$Ca$_2$SrCu$_2$O$_{8+\delta}$    & b & (\onlinecite{liu99})  & 40 & $2600\pm 300$ & $3800\pm 300$ & 0.418 \\
 Y-Bi$_2$Ca$_2$SrCu$_2$O$_{8+\delta}$    & b & (\onlinecite{wang99}) & 43 & $4200\pm 400$ & $5140\pm 500$ & 0.309 \\
 & & & & \\
 Tl$_2$Ba$_2$CuO$_{6+\delta}$ & b & (\onlinecite{puchkov95}) & 88  & $5000\pm 500$ & $6630\pm 500$ & 0.240 \\
% Tl$_2$Ba$_2$CuO$_{6+\delta}$         & 32  & $12\pm 3$    & $134\pm 15$   & \\
    & & & & & \\
%
% NCCO and PCCO, last to entries are film data...
%
 Nd$_{1.85}$Ce$_{0.15}$CuO$_4$ & b & (\onlinecite{homes97,singley01}) & 23   & $28\,000\pm 2000$ & $10\,300\pm 900$ & 0.154 \\
 Pr$_{1.85}$Ce$_{0.15}$CuO$_4$ & b & (\onlinecite{homes04b})  & 23 & $30\,000\pm 3000$ & $10\,300\pm 900$ & 0.154 \\
 Pr$_{1.85}$Ce$_{0.15}$CuO$_4$ & d & (\onlinecite{zimmers04}) & 21 & $15\,000\pm 2000$ &   $4820\pm  600$ & 0.330 \\
 Pr$_{1.87}$Ce$_{0.15}$CuO$_4$ & d & (\onlinecite{zimmers04}) & 16 & $50\,000\pm 6000$ &   $7960\pm  800$ & 0.207 \\
  & & & & \\
 La$_{1.87}$Sr$_{0.13}$CuO$_4$ & b & (\onlinecite{startseva99}) & 32 & $7000\pm 700$     & $5600\pm 450$  & 0.284 \\
 La$_{1.86}$Sr$_{0.14}$CuO$_4$ & b & (\onlinecite{startseva99}) & 36 & $9000\pm 900$     & $6000\pm 500$  & 0.265 \\
%
% BKBO data
%
  & & & & & \\
 Ba$_{0.62}$K$_{0.38}$BiO$_3$ & b & (\onlinecite{puchkov96b}) & 31 & $3800\pm 300$ & $4240\pm 400$ & 0.375 \\
 Ba$_{0.60}$K$_{0.40}$BiO$_3$ & b & (\onlinecite{puchkov96b}) & 28 & $4400\pm 300$ & $4240\pm 400$ & 0.375 \\
 Ba$_{0.54}$K$_{0.46}$BiO$_3$ & b & (\onlinecite{puchkov96b}) & 21 & $6000\pm 300$ & $4240\pm 400$ & 0.375 \\
 & & & & & \\
 Nb & d & (\onlinecite{pronin98}) & 8.3 & 2.5e5 & $17\,600$ & 0.090 \\
 Nb & d & (\onlinecite{klein94})  & 9.3 & 8.5e5 & $35\,800$ & 0.044 \\
 Pb & d & (\onlinecite{klein94})  & 7.2 & 1.4e6 & $41\,000$ & 0.038 \\
\end{tabular}
\end{ruledtabular}
\footnotetext[1] {Radiation damaged, twinned single crystal.}%
\footnotetext[2] {Light polarized in the {\it a-b} plane of a single crystal.}%
\footnotetext[3] {Light polarized along the {\it a} axis of a twin-free single crystal.}%
\footnotetext[4] {Thin film or oriented thin film.}%
\end{table*}
%
% end of squeezetable
%
%}
%
In general, BCS superconductors have relatively low values for $T_c$, thus
$1/\tau$ is assumed to have little temperature dependence close to the
superconducting transition. This assumption may be tested by suppressing $T_c$
through the application of a magnetic field in excess of the upper critical
field (H$_{c2}$) and examining the transport properties, which typically reveal
little temperature dependence in $1/\tau$ below the zero-field value of $T_c$.
The application of the the clean and dirty-limit picture to the copper-oxide
superconductors is complicated by both the high critical temperature, and the
superconducting energy gap which is thought to be {\it d}-wave in nature and
momentum dependent ($\Delta_k$), containing nodes.\cite{shen93,hardy93}  The
high value for $T_c$ suggests that $1/\tau$ may still have a significant
temperature dependence close to $T_c$.  In the normal state the scattering rate
is often observed to be rather large, scaling linearly with
temperature,\cite{iye92} and is presumed to be dominated by inelastic
processes. Indeed, below $T_c$ the quasiparticle scattering rate in the
cuprates is observed to decrease by nearly two orders of magnitude at low
temperatures.\cite{bonn93} This rapid decrease in $1/\tau$ is also observed
optically, but not to the same extent.\cite{puchkov96a} A gap with
$d_{x^2-y^2}$ symmetry may be written as $\Delta_k = \Delta_0 \left[ \cos(k_x
a) - \cos(k_y a) \right]$; the gap reaches a maximum at the $(0,\pi)$ and
$(\pi,0)$ points, and vanishes along the nodal $(\pi, \pi)$ directions.  The
fact that the scattering rate of the quasiparticles restricted to the nodal
regions of the Fermi surface for $T\ll T_c$ is quite small has been taken as
evidence that these materials are in the clean limit.\cite{sutherland03,
turner03,hill04} While it is certainly true that for $T\ll T_c$ the scattering
rate is small and the nodal quasiparticles have very long mean-free paths, it
is problematic to assert that the superconductor is therefore in the clean
limit. In a normal BCS superconductor $1/\tau$ is also observed to decrease
dramatically below $T_c$, regardless of the normal-state value of $1/\tau$, due
to the formation of a condensate.\cite{basov02} Thus, the criteria of a small
value of the quasiparticle scattering rate for $T\ll T_c$ is not necessarily a
good measure of whether or not the superconductivity is in the clean or dirty
limit.
%
% High-field results
%
As with BCS materials, it is desirable to suppress $T_c$ in the copper-oxide
materials through the application of a magnetic field to determine the
low-temperature behavior of $1/\tau$. While H$_{c2}$ is quite large in the cuprates,
experiments using pulsed magnetic fields can suppress $T_c$; in these experiments
the resistivity of the optimally-doped materials matches the zero-field values at
high temperatures due to the low magnetoresistance, and the trend of slowly
decreasing resistivity continues smoothly to low temperatures,\cite{boebinger96,
ando96,ando97a,ando97b,ono00} often saturating at a value close to that observed at
$T_c$.  The implication of these experiments is that the normal-state value of
$1/\tau$ is a good measure of the scattering rate in those systems in which $T_c$
has been suppressed, and is therefore the value that should be considered when
determining whether a system is in the clean or dirty limit. In addition to this
explicit approach, a simpler method is to adopt an operational definition which
states that if $\rho_s$ changes with respect to the normal-state value of $1/\tau$
then the material is in the dirty limit; when $\rho_s$ is no longer sensitive to the
value of $1/\tau$ then the material is in the clean limit.
%
% Note about nature of data points...
%
Most of the materials in Fig.~\ref{fig:abplane} are studied as a function of carrier
doping, but it is also important to note that the introduction of disorder for fixed
doping levels has also been studied.\cite{basov94}  The fact that all the observed
results follow this linear scaling relation strongly suggests that many of the
copper-oxide superconductors are close to or in the dirty limit (i.e., the
superfluid density changes in response to variations in $1/\tau$).

%
% The clean limit.
%
\subsection{Clean limit}
\begin{figure}[t]
%
% eprint
%
%\vspace*{-0.5cm}%
%\centerline{\includegraphics[width=3.8in]{figure3.eps}}%
%\centerline{\includegraphics[width=3.8in]{figure3.eps}}%
%\vspace*{-0.3cm}%
%
% manuscript
%
\centerline{\includegraphics[width=3.2in]{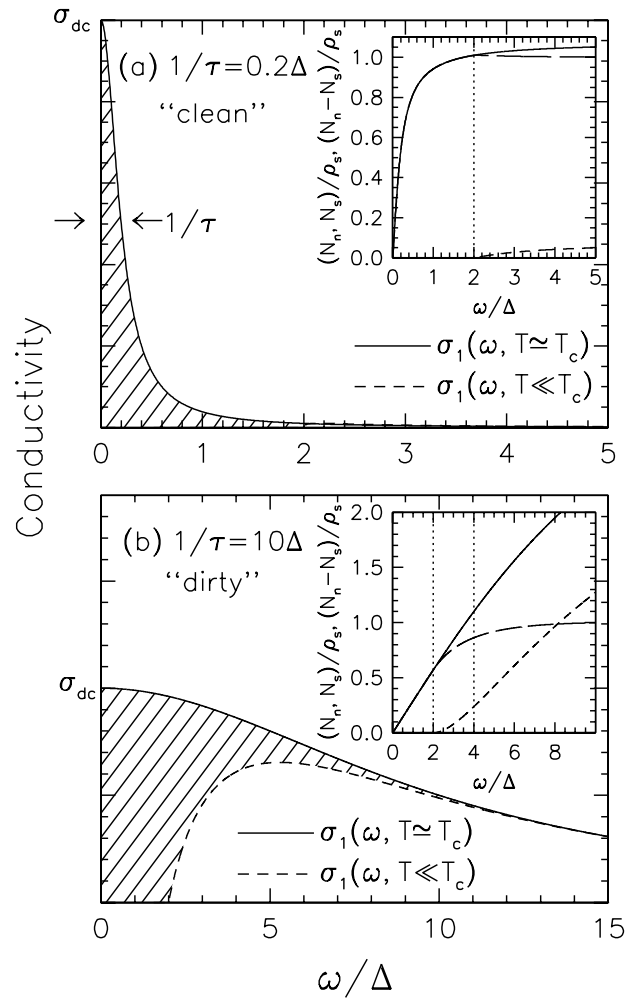}}%
%\centerline{\includegraphics[width=3.8in]{figure3.eps}}%
%\vspace*{-0.3cm}%
%
\caption{The optical conductivity for the BCS model in the normal (solid line)
and superconducting states (dashed line) for a material in (a) the clean limit
($1/\tau \ll 2\Delta$), and (b) the dirty limit ($1/\tau > 2\Delta$). The
normal-state conductivity is a Lorentzian centered at zero frequency with a
full width at half maximum of $1/\tau$ for $T\simeq T_c$.  The spectral weight
associated with the formation of a superconducting condensate is indicated by
the hatched area.
Insets: $N_n = N(\omega, T\simeq T_c)/\rho_s$ (solid line),  $N_s = N(\omega, T
\ll T_c)/\rho_s$ (dashed line), and difference between the two (long-dashed
line), normalized with respect to $\rho_s$; in the clean limit $(N_n -
N_s)/\rho_s$ converges rapidly to unity, and is fully formed at energies
comparable to $1/\tau$, while in the dirty limit convergence occurs at energies
comparable to $4\Delta$.}
\vspace*{-0.2cm}%
\label{fig:model}
\end{figure}

The BCS model is used to describe the superconductivity of simple metals and alloys.
If the normal-state properties may be described by the simple Drude model in which
the complex dielectric function is written as $\tilde\epsilon(\omega) =
\epsilon_\infty - \omega_p^2/[\omega(\omega + i\gamma)]$, where $\omega_p^2=4\pi
ne^2/m^\ast$ is the classical plasma frequency with the free-carrier concentration
$n$ and effective mass $m^\ast$, $\gamma=1/\tau$ is the scattering rate, and
$\epsilon_\infty$ is a high-frequency contribution. The dielectric function and the
conductivity are related through $\tilde\sigma = \sigma_1+i\sigma_2 =
-i\omega\tilde\epsilon/ 4\pi$, thus the real part of the frequency-dependent
conductivity has the form $\sigma_1(\omega) = \sigma_{dc}/(1+\omega^2\tau^2)$ and
$\sigma_{dc} = \omega_p^2\tau/4\pi$, which has the shape of a Lorentzian centered at
zero frequency with a width at half-maximum given by $1/\tau$. The optical
conductivity below $T_c$ has been calculated from an isotropic ({\it s}-wave) energy
gap $2\Delta$ that considers an arbitrary purity level.\cite{zimmerman91} The clean
limit case ($1/\tau \ll 2\Delta$) is illustrated in Fig.~\ref{fig:model}(a) for the
choice $1/\tau = 0.2\Delta$.
An aspect of clean-limit systems is that nearly all of the spectral weight
associated with the condensate lies below $2\Delta$. As a result, the
normalized spectral weight of the condensate\cite{basov99} $(N_n-N_s)/\rho_s$
shown in the inset of Fig.~\ref{fig:model}(a) approaches unity at frequencies
closer to $1/\tau$ rather than $2\Delta$.  The spectral weight for the
condensate (the difference in the area under the two curves, indicated by the
hatched region) may be estimated as $\rho_s \simeq \sigma_{dc}/\tau$. If
$1/\tau \propto T_c$ for $T\simeq T_c$ in the copper-oxide
materials,\cite{orenstein90} then $\rho_s \propto \sigma_{dc}\,T_c$, in
agreement with the observed scaling relation.  It is interesting to note that
$1/\tau \propto T_c$ yields rather large values for the normal-state scattering
rate, and it has been suggested that the copper-oxide materials are close to
the maximum level of dissipation allowed for these systems.\cite{zaanen04}
Furthermore, even though a {\it d}-wave system complicates the interpretation
of the clean and dirty limit, large normal-state values of $1/\tau$ and
relatively short normal-state mean-free paths\cite{lee04} are problematic for a
clean-limit picture; to achieve the clean limit it is not only necessary that
$1/\tau \ll 2\Delta_0$, but also that $1/\tau \lesssim 2\Delta_k$ in the nodal
regions.  In fact, the clean-limit requirement is much more stringent for a
{\it d}-wave system than it is for a material with an isotropic energy gap, and
it is not clear that it will ever be satisfied in the copper-oxide
superconductors.  This suggests that a dirty-limit view may be more
appropriate.

%
% Figure 3 - clean and dirty limit spectral weight
%

%
% The dirty limit.
%
\subsection{Dirty limit}
In the BCS dirty limit, $1/\tau > 2\Delta$; this is illustrated in
Fig.~\ref{fig:model}(b) for $1/\tau = 10\Delta$.  In this case the normal-state
conductivity is a considerably broader Lorentzian, and much of the spectral
weight has been pushed out above $2\Delta$.  As a result, the normalized
spectral weight of the condensate, shown in the inset, converges much more
slowly than in the clean-limit case. However, a majority of the spectral weight
is captured by $2\Delta$ and $\rho_s$ is almost fully formed above $4\Delta$
(Ref.~\onlinecite{basov99}). In the dirty-limit case, the spectral weight of
the condensate [the hatched area in Fig.~\ref{fig:model}(a)] may be estimated
as $\rho_s \simeq \sigma_{dc}\,2\Delta$. In the BCS model, the energy gap
$2\Delta$ scales linearly with $T_c$, yielding $\rho_s \propto
\sigma_{dc}\,T_c$, which is in agreement with the observed scaling relation. As
in the clean-limit case, the nature of the gap is important.  If $1/\tau >
2\Delta_0$, the spirit of the dirty-limit case is preserved for all $\Delta_k$.
While many of the points in Figs.~\ref{fig:linear} \& \ref{fig:abplane} are
doping-dependent studies and do not track systematic changes in $1/\tau$, some
of these points are for the same chemical doping with different scattering
rates resulting from disorder that have either been deliberately
introduced,\cite{basov94} or that exist simply as a byproduct of synthesis
(Table~I).\cite{hardy94,hosseini99} The observation that all the points obey a
linear scaling relation satisfies the operational definition of the dirty
limit, suggesting that the examined materials are either close to or in the
dirty limit.

%
% Figure 4
%
\begin{figure}[t]
%
% manuscript
%
%\vspace*{-0.5cm}%
%\centerline{\includegraphics[width=3.8in]{figur4.eps}}%
%\centerline{\includegraphics[width=6.0in]{figure4.eps}}%
%\vspace*{-0.3cm}%
%
\centerline{\includegraphics[width=3.4in]{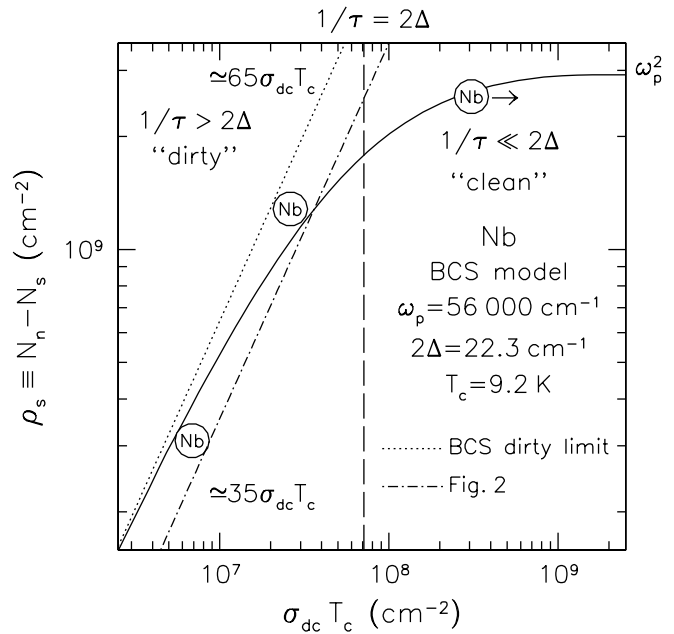}}%
\vspace*{-0.3cm}%
\caption{The log-log plot of the predicted behavior from the BCS model of the
strength of the condensate for Nb for a wide range of scattering rates $1/\tau =
0.05\Delta \rightarrow 50\Delta$, and assuming a plasma frequency $\omega_p =
56\,000$~cm$^{-1}$, critical temperature $T_c = 9.2$~K and an energy gap of $2\Delta
= 3.5\,k_B\,T_c$ (solid line).  The dashed line indicates $1/\tau = 2\Delta$.  To
the right of this line the material approaches the clean limit with a residual
resistance ratio (RRR) of $\gtrsim 100$; the right arrow indicates that for larger
RRR's, $\sigma_{dc}$ close to $T_c$ increases, but $\rho_s$ has saturated to
$\omega_p^2$ (the data point for Nb in this regime is from
Ref.~\onlinecite{varmazis74}). As the scattering rate increases, the strength of the
condensate adopts a linear scaling behavior (dotted line); the two points for Nb
(Refs.~\onlinecite{klein94,pronin98}) shown in Fig.~\ref{fig:abplane} lie close to
this line, indicating that they are in the dirty limit.  The scaling relation shown
in Fig.~\ref{fig:abplane}
(dash-dot line) is slightly offset from the BCS dirty-limit result. }%
\vspace*{-0.2cm}%
\label{fig:nb}
\end{figure}

%
% Discussion of Nb
%
\subsection{Behavior of Nb}
It was noted in Fig.~\ref{fig:abplane} that the points for Nb and Pb agreed
reasonably well with the scaling relation used to describe the copper-oxide
superconductors.  It is important to determine if these values represent clean
or dirty-limit results.  The expected behavior of Nb has been modeled using the
BCS model\cite{zimmerman91} for an arbitrary purity level with a critical
temperature of $T_c = 9.2$~K and a gap of $2\Delta = 22.3$~cm$^{-1}$ (the BCS
weak-coupling limit $2\Delta = 3.5\,k_B\,T_c$). The normal-state is described
using the Drude model with a classical plasma frequency of $\omega_p =
56\,000$~cm$^{-1}$ (Ref.~\onlinecite{palik}) and a range of scattering rates
$1/\tau = 0.05\Delta \rightarrow 50\Delta$; from the Drude model the dc
conductivity is $\sigma_{dc} = \omega_p^2\,\tau/60$ (in units of
$\Omega^{-1}$cm$^{-1}$ when the plasma frequency and the scattering rate have
units of cm$^{-1}$). The spectral weight of the condensate $\rho_s = N_n - N_s$
has been determined by integrating to $\omega \simeq 200\Delta$, where $\rho_s$
is observed to converge for all the values of $1/\tau$ examined. The result of
this calculation is shown as the solid line in Fig.~\ref{fig:nb}, and the
vertical dashed line indicates where $1/\tau = 2\Delta$. The point to the right
of the dashed line is for Nb recrystallized in ultra-high
vacuum\cite{varmazis74} to achieve clean-limit conditions in which the residual
resistivity ratios [$\rho({\rm RT}) / \rho(T\gtrsim T_c)$] are well in excess
of 100, and where $\rho_s \rightarrow \omega_p^2$ for $T \ll T_c$. As the
scattering rate increases and the material becomes progressively more
``dirty'', the strength of the condensate begins to decrease until it adopts
the linear scaling behavior $\rho_s \simeq 65\,\sigma_{dc}\,T_c$ observed in
Fig.~\ref{fig:nb}. (It should be noted that the BCS model yields the same
asymptotic behavior in the dirty limit, regardless of the choice of $\omega_p$
or $\Delta$; the constant is only sensitive upon the ratio of $\Delta$ to
$T_c$.) The two points for Nb shown in Fig.~\ref{fig:abplane}, (reproduced in
Fig.~\ref{fig:nb}), fall close to this line\cite{klein94,pronin98} and are
clearly in the dirty limit. Thus, the scaling relation $\rho_s \propto
\sigma_{dc}\,T_c$ is the hallmark of a BCS dirty-limit system.\cite{smith92}
%
% Relation to the high-Tc materials...
%
The scaling relation for the copper-oxide superconductors $\rho_s \simeq
35\,\sigma_{dc}\,T_c$ is somewhat less than the $\rho_s \simeq
65\,\sigma_{dc}\,T_c$ asymptotic behavior observed for the weak-limit BCS
material.  However, in the log-log representation of Fig.~\ref{fig:abplane},
the numerical constant in the scaling relation is the offset of the line.  The
line may be shifted by assuming different ratios between $2\Delta$ and
$k_BT_c$; the initial value of $\simeq 65$ was based on the weak-coupling value
of $2\Delta/k_B\,T_c \simeq 3.5$, while the observed value of $\simeq 35$ may
be reproduced by assuming a smaller ratio $2\Delta/k_B\,T_c \simeq 2$.  The
difference may also arise from the different symmetry of the superconducting
energy gap in the two systems, and the fact that in the copper-oxide materials
there is still a substantial amount of low-frequency residual conductivity at
low temperature.
Regardless of these differences, the empirical scaling relation $\rho_s \propto
\sigma_{dc}\,T_c$ is observed in both the copper oxide and disordered elemental
superconductors. If it is true in general that $\rho_s \propto
\sigma_{dc}\,2\Delta$, then this necessarily implies that $\Delta \propto T_c$.
In the optimally-doped and overdoped materials, there is some evidence that
$\Delta_0 \propto T_c$ (Refs.~\onlinecite{panagopoulos98,damascelli03}).
In the underdoped materials, large gaps are observed to develop in the normal
state\cite{yeh01} well above $T_c$.  While it has been noted that the energy
scale over which spectral weight is transferred into the condensate is much
larger in the underdoped materials than it is for the optimally-doped
materials,\cite{syro02,homes04a} the majority of the spectral weight is still
captured at energies comparable to $T_c$.  This would tend to support the view
that the energy scale relevant to phase coherence and the formation of the
condensate is proportional to $T_c$.

%
% Link to phenomenological approach
%
It is of some interest at this point to compare the empirical relation, that
$\rho_s$ is proportional to $\sigma_{dc}\,T_c$, with the expression for the
penetration depth that is given by the Ginzburg-Landau theory modified for the
dirty limit. In general, the expression for the London penetration depth is
given by $\lambda_L(T\rightarrow 0) =  \sqrt{m c^2/(4\pi n_s e^2)}$, where $n_s
\equiv n$ is the superconducting carrier concentration.  In the dirty limit one
can show that $\rho_s({\rm dirty})/\rho_s({\rm clean}) = l/\xi_0$
(Ref.~\onlinecite{ferrell67}). An increase in $1/\tau$ reduces the amount of
superfluid and the penetration depth increases and can be written as $\lambda^2
= (\xi_0/l)\,\lambda_L^2$. Since $\lambda^2 \propto 1/\rho_s$, $\xi_0 \propto
1/T_c$, and $\sigma_{dc} \propto l$, then one can recover the result that
$\rho_s \propto \sigma_{dc}\,T_c$.  It is possible that in a {\it d}-wave
system the presence of nodal regions with a small superfluid density and
$\Delta_k \ll \Delta_0$, that the coherence length in the above expression for
$\lambda^2$ now involves some average including the nodal regions.

%
% Figure 5
%
\begin{figure}[t]
%
% manuscript
%
%\vspace*{-0.5cm}%
%\centerline{\includegraphics[width=3.8in]{figur5.eps}}%
%\centerline{\includegraphics[width=6.0in]{figure5.eps}}%
%\vspace*{-0.3cm}%
%
\centerline{\includegraphics[width=3.4in]{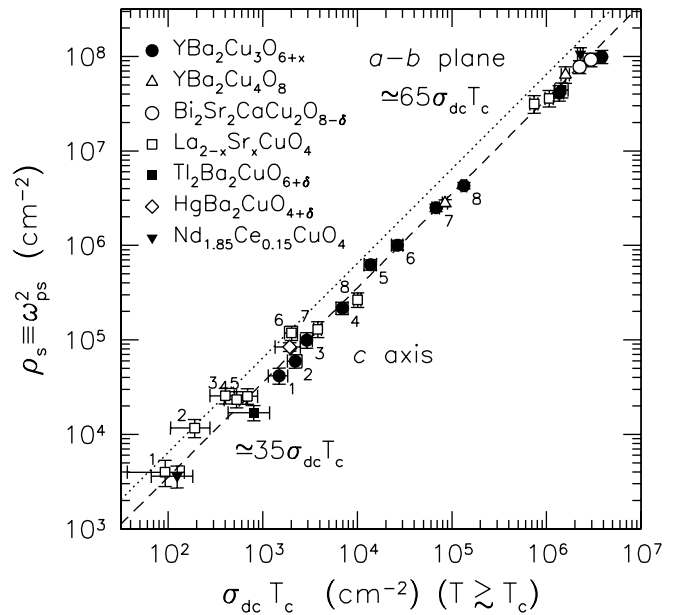}}%
\vspace*{-0.3cm}%
\caption{The log-log plot of the superfluid density expressed as a plasma frequency
$\rho_s \equiv \omega_{pS}^2$ vs $\sigma_{dc}\,T_c$ for the {\it a-b} planes and the
{\it c} axis for a variety of cuprates. Within error, all of the points fall on the
same universal (dashed) line  defined by $\rho_s \simeq 35\,\sigma_{dc}\,T_c$; the
dotted line is the dirty limit result $\rho_s \simeq 65\,\sigma_{dc}\,T_c$ for the
BCS weak-coupling case ($2\Delta =3.5\,k_B\,T_c$) from Fig.~\ref{fig:nb}, and also
represents the Josephson result for the BCS weak-coupling case, used to describe the
scaling along the {\it c}-axis.\cite{homes03}  (Values for the {\it c}-axis points
are listed in the supplemental infomation of Ref.~\onlinecite{homes04b}.)}%
\vspace*{-0.3cm}%
\label{fig:total}
\end{figure}

%
% The c axis.
%
\subsection{The c axis}
It was previously observed\cite{homes04b} that the scaling relation $\rho_s
\simeq 35\sigma_{dc}\,T_c$ is a universal result that describes not only the
the {\it a-b} planes, but the {\it c} axis as well, as shown in
Fig.~\ref{fig:total}. While a description of the scaling based scattering rates
within the context of clean and dirty limits may be appropriate for the {\it
a-b} planes where the transport is coherent, it is inappropriate along the {\it
c} axis, where the activated nature of the temperature dependence of the
resistivity indicates that the transport in this direction is incoherent and
governed by hopping.\cite{cooper94} In this case, the superconductivity along
the {\it c} axis may be described by the Josephson effect, which for the BCS
weak coupling case ($2\Delta = 3.5\,k_B\,T_c$) yields $\rho_s \simeq
65\,\sigma_{dc}\,T_c$ (Ref.~\onlinecite{homes03}). Surprisingly, this is
precisely the result that was obtained in the {\it a-b} planes for the BCS
weak-coupling case in the dirty limit in Fig.~\ref{fig:nb}, indicating that
from a functional point of view the dirty limit and the Josephson effect are
nearly identical.  One interpretation of this result is that the Josephson
effect may arise naturally out of systems with an increasing amount of
disorder, and that as a result any crossover from coherent to incoherent
behavior results in the same overall scaling relation.  Another somewhat more
speculative possibility is that the copper-oxide superconductors may be so
electronically inhomogeneous that it may be possible to view the Josephson
effect as appropriate not only for the {\it c} axis, but for the {\it a-b}
planes as well.\cite{jung98,tu04}

%
% Summary...
%
\section{Conclusions}
The implications of the linear scaling relation $\rho_s \propto
\sigma_{dc}\,T_c$ in the copper-oxide superconductors has been examined within
the context of clean and dirty-limit systems.  In the conventional BCS
superconductors (such as Nb), this linear scaling is the hallmark of a
dirty-limit system.  The copper-oxide materials are thought to be {\it d}-wave
superconductors, in which the clean limit is difficult to achieve. The observed
linear scaling strongly suggests that the copper-oxide superconductors are
either close to or in the dirty limit. Estimates of $\rho_s$ based on geometric
arguments imply that the energy scale below which the majority of the spectral
weight is transferred into the condensate scales linearly with $T_c$.  The {\it
a-b} planes and the {\it c} axis follow the same scaling
relation.\cite{homes04b} The scaling behavior for the dirty limit and the
Josephson effect (assuming a BCS formalism) is essentially identical from a
functional point of view, suggesting that in some regime the dirty limit and
the Josephson effect may be viewed as equivalent.  This raises the possibility
that electronic inhomogeneities may play an important role in the mechanism of
superconductivity in the copper-oxide high-temperature superconductors.

%
% Acknowledgements
%
\begin{acknowledgments}
The authors would like to thank Y.~Ando, D.~N.~Basov, D.~A.~Bonn, I.~Bozovic,
A.~V.~Chubukov, M.~Greven, W.~N.~Hardy, P.~D.~Johnson, S.~A.~Kivelson, P.~A.~Lee,
T.~M.~Rice, D.~B.~Tanner, T.~Timusk, and J.~J.~Tu for useful discussions. Work at
Brookhaven was supported by the DOE under contract number DE-AC02-98CH10886.
\end{acknowledgments}

%
%%%%%%%%%%%%%%%%%%%%%%%%%%%%%%%%%%%%%%%%%%%%%%%%%%%%%%%%%%%%%%%%%%%%%%%%%%%%%%
%
% References
%
\bibliography{abplane}

\begin{thebibliography}{65}
\expandafter\ifx\csname natexlab\endcsname\relax\def\natexlab#1{#1}\fi
\expandafter\ifx\csname bibnamefont\endcsname\relax
  \def\bibnamefont#1{#1}\fi
\expandafter\ifx\csname bibfnamefont\endcsname\relax
  \def\bibfnamefont#1{#1}\fi
\expandafter\ifx\csname citenamefont\endcsname\relax
  \def\citenamefont#1{#1}\fi
\expandafter\ifx\csname url\endcsname\relax
  \def\url#1{\texttt{#1}}\fi
\expandafter\ifx\csname urlprefix\endcsname\relax\def\urlprefix{URL }\fi
\providecommand{\bibinfo}[2]{#2}
\providecommand{\eprint}[2][]{\url{#2}}

\bibitem[{\citenamefont{Whittaker}(1999)}]{whittaker99}
\bibinfo{author}{\bibfnamefont{R.~J.} \bibnamefont{Whittaker}},
  \bibinfo{journal}{Nature} \textbf{\bibinfo{volume}{401}},
  \bibinfo{pages}{865} (\bibinfo{year}{1999}).

\bibitem[{\citenamefont{Brown and West}(1999)}]{brown99}
\bibinfo{editor}{\bibfnamefont{J.~H.} \bibnamefont{Brown}} \bibnamefont{and}
  \bibinfo{editor}{\bibfnamefont{G.~B.} \bibnamefont{West}}, eds.,
  \emph{\bibinfo{title}{Scaling in biology}} (\bibinfo{publisher}{Oxford
  University Press}, \bibinfo{address}{Oxford}, \bibinfo{year}{1999}).

\bibitem[{\citenamefont{Bednorz and Mueller}(1986)}]{bednorz86}
\bibinfo{author}{\bibfnamefont{J.~G.} \bibnamefont{Bednorz}} \bibnamefont{and}
  \bibinfo{author}{\bibfnamefont{K.~A.} \bibnamefont{Mueller}},
  \bibinfo{journal}{Z. Phys. B} \textbf{\bibinfo{volume}{64}},
  \bibinfo{pages}{189} (\bibinfo{year}{1986}).

\bibitem[{\citenamefont{Schneider}(2002)}]{schneider02}
\bibinfo{author}{\bibfnamefont{T.}~\bibnamefont{Schneider}},
  \bibinfo{journal}{Europhys. Lett.} \textbf{\bibinfo{volume}{60}},
  \bibinfo{pages}{141} (\bibinfo{year}{2002}).

\bibitem[{\citenamefont{Uemura et~al.}(1989)\citenamefont{Uemura, Luke,
  Sternlieb, Brewer, Carolan, Hardy, Kadono, Kiefl, Kreitzman, Mulhern
  et~al.}}]{uemura89}
\bibinfo{author}{\bibfnamefont{Y.~J.} \bibnamefont{Uemura}},
  \bibinfo{author}{\bibfnamefont{G.~M.} \bibnamefont{Luke}},
  \bibinfo{author}{\bibfnamefont{B.~J.} \bibnamefont{Sternlieb}},
  \bibinfo{author}{\bibfnamefont{J.~H.} \bibnamefont{Brewer}},
  \bibinfo{author}{\bibfnamefont{J.~F.} \bibnamefont{Carolan}},
  \bibinfo{author}{\bibfnamefont{W.~N.} \bibnamefont{Hardy}},
  \bibinfo{author}{\bibfnamefont{R.}~\bibnamefont{Kadono}},
  \bibinfo{author}{\bibfnamefont{R.~F.} \bibnamefont{Kiefl}},
  \bibinfo{author}{\bibfnamefont{S.~R.} \bibnamefont{Kreitzman}},
  \bibinfo{author}{\bibfnamefont{P.}~\bibnamefont{Mulhern}},
  \bibnamefont{et~al.}, \bibinfo{journal}{Phys. Rev. Lett.}
  \textbf{\bibinfo{volume}{62}}, \bibinfo{pages}{2317} (\bibinfo{year}{1989}).

\bibitem[{\citenamefont{Uemura et~al.}(1991)\citenamefont{Uemura, Le, Luke,
  Sternlieb, Wu, Brewer, Riseman, Seaman, Maple, Ishikawa et~al.}}]{uemura91}
\bibinfo{author}{\bibfnamefont{Y.~J.} \bibnamefont{Uemura}},
  \bibinfo{author}{\bibfnamefont{L.~P.} \bibnamefont{Le}},
  \bibinfo{author}{\bibfnamefont{G.~M.} \bibnamefont{Luke}},
  \bibinfo{author}{\bibfnamefont{B.~J.} \bibnamefont{Sternlieb}},
  \bibinfo{author}{\bibfnamefont{W.~D.} \bibnamefont{Wu}},
  \bibinfo{author}{\bibfnamefont{J.~H.} \bibnamefont{Brewer}},
  \bibinfo{author}{\bibfnamefont{T.~M.} \bibnamefont{Riseman}},
  \bibinfo{author}{\bibfnamefont{C.~L.} \bibnamefont{Seaman}},
  \bibinfo{author}{\bibfnamefont{M.~B.} \bibnamefont{Maple}},
  \bibinfo{author}{\bibfnamefont{M.}~\bibnamefont{Ishikawa}},
  \bibnamefont{et~al.}, \bibinfo{journal}{Phys. Rev. Lett.}
  \textbf{\bibinfo{volume}{66}}, \bibinfo{pages}{2665} (\bibinfo{year}{1991}).

\bibitem[{\citenamefont{Zuev et~al.}()\citenamefont{Zuev, Kim, and
  Lemberger}}]{zuev04}
\bibinfo{author}{\bibfnamefont{Y.}~\bibnamefont{Zuev}},
  \bibinfo{author}{\bibfnamefont{M.~S.} \bibnamefont{Kim}}, \bibnamefont{and}
  \bibinfo{author}{\bibfnamefont{T.~R.} \bibnamefont{Lemberger}},
  \eprint{cond-mat/0410135}.

\bibitem[{\citenamefont{Niedermayer et~al.}(1993)\citenamefont{Niedermayer,
  Bernhard, Binninger, Glückler, Tallon, Ansaldo, and Budnick}}]{niedermayer93}
\bibinfo{author}{\bibfnamefont{C.}~\bibnamefont{Niedermayer}},
  \bibinfo{author}{\bibfnamefont{C.}~\bibnamefont{Bernhard}},
  \bibinfo{author}{\bibfnamefont{U.}~\bibnamefont{Binninger}},
  \bibinfo{author}{\bibfnamefont{H.}~\bibnamefont{Glückler}},
  \bibinfo{author}{\bibfnamefont{J.~L.} \bibnamefont{Tallon}},
  \bibinfo{author}{\bibfnamefont{E.~J.} \bibnamefont{Ansaldo}},
  \bibnamefont{and} \bibinfo{author}{\bibfnamefont{J.~I.}
  \bibnamefont{Budnick}}, \bibinfo{journal}{Phys. Rev. Lett.}
  \textbf{\bibinfo{volume}{71}}, \bibinfo{pages}{1764–1767}
  (\bibinfo{year}{1993}).

\bibitem[{\citenamefont{Tallon et~al.}(2003)\citenamefont{Tallon, Loram,
  Cooper, Panagopoulos, and Bernhard}}]{tallon03}
\bibinfo{author}{\bibfnamefont{J.~L.} \bibnamefont{Tallon}},
  \bibinfo{author}{\bibfnamefont{J.~W.} \bibnamefont{Loram}},
  \bibinfo{author}{\bibfnamefont{J.~R.} \bibnamefont{Cooper}},
  \bibinfo{author}{\bibfnamefont{C.}~\bibnamefont{Panagopoulos}},
  \bibnamefont{and} \bibinfo{author}{\bibfnamefont{C.}~\bibnamefont{Bernhard}},
  \textbf{\bibinfo{volume}{68}}, \bibinfo{pages}{180501}
  (\bibinfo{year}{2003}).

\bibitem[{\citenamefont{Homes et~al.}(1997)\citenamefont{Homes, Clayman, Peng,
  and Greene}}]{homes97}
\bibinfo{author}{\bibfnamefont{C.~C.} \bibnamefont{Homes}},
  \bibinfo{author}{\bibfnamefont{B.~P.} \bibnamefont{Clayman}},
  \bibinfo{author}{\bibfnamefont{J.~L.} \bibnamefont{Peng}}, \bibnamefont{and}
  \bibinfo{author}{\bibfnamefont{R.~L.} \bibnamefont{Greene}},
  \bibinfo{journal}{Phys. Rev. B} \textbf{\bibinfo{volume}{56}},
  \bibinfo{pages}{5525} (\bibinfo{year}{1997}).

\bibitem[{\citenamefont{Homes et~al.}(1998)\citenamefont{Homes, Clayman, Peng,
  and Greene}}]{homes98a}
\bibinfo{author}{\bibfnamefont{C.~C.} \bibnamefont{Homes}},
  \bibinfo{author}{\bibfnamefont{B.~P.} \bibnamefont{Clayman}},
  \bibinfo{author}{\bibfnamefont{J.~L.} \bibnamefont{Peng}}, \bibnamefont{and}
  \bibinfo{author}{\bibfnamefont{R.~L.} \bibnamefont{Greene}},
  \bibinfo{journal}{J. Phys. Chem. Solids} \textbf{\bibinfo{volume}{59}},
  \bibinfo{pages}{1979} (\bibinfo{year}{1998}).

\bibitem[{\citenamefont{Pimenov et~al.}(1999)\citenamefont{Pimenov, Loidl,
  Schey, Stritzker, Jakob, Adrian, Pronin, and Goncharov}}]{pimenov99}
\bibinfo{author}{\bibfnamefont{A.}~\bibnamefont{Pimenov}},
  \bibinfo{author}{\bibfnamefont{A.}~\bibnamefont{Loidl}},
  \bibinfo{author}{\bibfnamefont{B.}~\bibnamefont{Schey}},
  \bibinfo{author}{\bibfnamefont{B.}~\bibnamefont{Stritzker}},
  \bibinfo{author}{\bibfnamefont{G.}~\bibnamefont{Jakob}},
  \bibinfo{author}{\bibfnamefont{H.}~\bibnamefont{Adrian}},
  \bibinfo{author}{\bibfnamefont{A.~V.} \bibnamefont{Pronin}},
  \bibnamefont{and} \bibinfo{author}{\bibfnamefont{Y.~G.}
  \bibnamefont{Goncharov}}, \bibinfo{journal}{Europhys. Lett.}
  \textbf{\bibinfo{volume}{48}}, \bibinfo{pages}{73} (\bibinfo{year}{1999}).

\bibitem[{\citenamefont{Homes et~al.}(2004{\natexlab{a}})\citenamefont{Homes,
  Dordevic, Strongin, Bonn, Liang, Hardy, Komiya, Ando, Yu, Kaneko
  et~al.}}]{homes04b}
\bibinfo{author}{\bibfnamefont{C.~C.} \bibnamefont{Homes}},
  \bibinfo{author}{\bibfnamefont{S.~V.} \bibnamefont{Dordevic}},
  \bibinfo{author}{\bibfnamefont{M.}~\bibnamefont{Strongin}},
  \bibinfo{author}{\bibfnamefont{D.~A.} \bibnamefont{Bonn}},
  \bibinfo{author}{\bibfnamefont{R.}~\bibnamefont{Liang}},
  \bibinfo{author}{\bibfnamefont{W.~N.} \bibnamefont{Hardy}},
  \bibinfo{author}{\bibfnamefont{S.}~\bibnamefont{Komiya}},
  \bibinfo{author}{\bibfnamefont{Y.}~\bibnamefont{Ando}},
  \bibinfo{author}{\bibfnamefont{G.}~\bibnamefont{Yu}},
  \bibinfo{author}{\bibfnamefont{N.}~\bibnamefont{Kaneko}},
  \bibnamefont{et~al.}, \bibinfo{journal}{Nature}
  \textbf{\bibinfo{volume}{430}}, \bibinfo{pages}{539}
  (\bibinfo{year}{2004}{\natexlab{a}}).

\bibitem[{\citenamefont{Tallon et~al.}()\citenamefont{Tallon, Cooper, Naqib,
  and Loram}}]{tallon04}
\bibinfo{author}{\bibfnamefont{J.~L.} \bibnamefont{Tallon}},
  \bibinfo{author}{\bibfnamefont{J.~R.} \bibnamefont{Cooper}},
  \bibinfo{author}{\bibfnamefont{S.~H.} \bibnamefont{Naqib}}, \bibnamefont{and}
  \bibinfo{author}{\bibfnamefont{J.~W.} \bibnamefont{Loram}},
  \eprint{cond-mat/0410568}.

\bibitem[{\citenamefont{Homes et~al.}(1993)\citenamefont{Homes, Reedyk,
  Crandles, and Timusk}}]{homes93}
\bibinfo{author}{\bibfnamefont{C.~C.} \bibnamefont{Homes}},
  \bibinfo{author}{\bibfnamefont{M.}~\bibnamefont{Reedyk}},
  \bibinfo{author}{\bibfnamefont{D.}~\bibnamefont{Crandles}}, \bibnamefont{and}
  \bibinfo{author}{\bibfnamefont{T.}~\bibnamefont{Timusk}},
  \bibinfo{journal}{Appl. Opt.} \textbf{\bibinfo{volume}{32}},
  \bibinfo{pages}{2972} (\bibinfo{year}{1993}).

\bibitem[{uni()}]{units}
\bibinfo{note}{The term $120/\pi$ before the integral is the result of both a
  normalization of the integral (a geometric term) and a conversion of units,
  where it is assumed that the units of conductivity are in
  $\Omega^{-1}$cm$^{-1}$, and that the frequency is in cm$^{-1}$, so that the
  integral yields cm$^{-2}$. To recast all the units into cm$^{-1}$, two other
  unit conversions are implicitly employed in the text; 1~cm$^{-1} =
  0.21$~$\Omega^{-1}$cm$^{-1}$ and 1~cm$^{-1} = 1.44$~K.}

\bibitem[{\citenamefont{Ferrell and Glover{, }III}(1958)}]{ferrell58}
\bibinfo{author}{\bibfnamefont{R.~A.} \bibnamefont{Ferrell}} \bibnamefont{and}
  \bibinfo{author}{\bibfnamefont{R.~E.} \bibnamefont{Glover{, }III}},
  \bibinfo{journal}{Phys. Rev.} \textbf{\bibinfo{volume}{109}},
  \bibinfo{pages}{1398} (\bibinfo{year}{1958}).

\bibitem[{\citenamefont{Tinkham and Ferrell}(1959)}]{tinkham59}
\bibinfo{author}{\bibfnamefont{M.}~\bibnamefont{Tinkham}} \bibnamefont{and}
  \bibinfo{author}{\bibfnamefont{R.~A.} \bibnamefont{Ferrell}},
  \bibinfo{journal}{Phys. Rev. Lett.} \textbf{\bibinfo{volume}{2}},
  \bibinfo{pages}{331} (\bibinfo{year}{1959}).

\bibitem[{\citenamefont{Basov et~al.}(1999)\citenamefont{Basov, Woods, Katz,
  Singley, Dynes, Xu, Hinks, Homes, and Strongin}}]{basov99}
\bibinfo{author}{\bibfnamefont{D.~N.} \bibnamefont{Basov}},
  \bibinfo{author}{\bibfnamefont{S.~I.} \bibnamefont{Woods}},
  \bibinfo{author}{\bibfnamefont{A.~S.} \bibnamefont{Katz}},
  \bibinfo{author}{\bibfnamefont{E.~J.} \bibnamefont{Singley}},
  \bibinfo{author}{\bibfnamefont{R.~C.} \bibnamefont{Dynes}},
  \bibinfo{author}{\bibfnamefont{M.}~\bibnamefont{Xu}},
  \bibinfo{author}{\bibfnamefont{D.~G.} \bibnamefont{Hinks}},
  \bibinfo{author}{\bibfnamefont{C.~C.} \bibnamefont{Homes}}, \bibnamefont{and}
  \bibinfo{author}{\bibfnamefont{M.}~\bibnamefont{Strongin}},
  \bibinfo{journal}{Science} \textbf{\bibinfo{volume}{283}},
  \bibinfo{pages}{49} (\bibinfo{year}{1999}).

\bibitem[{\citenamefont{Basov et~al.}(1994)\citenamefont{Basov, Puchkov,
  Hughes, Strach, Preston, Timusk, Bonn, Liang, and Hardy}}]{basov94}
\bibinfo{author}{\bibfnamefont{D.~N.} \bibnamefont{Basov}},
  \bibinfo{author}{\bibfnamefont{A.~V.} \bibnamefont{Puchkov}},
  \bibinfo{author}{\bibfnamefont{R.~A.} \bibnamefont{Hughes}},
  \bibinfo{author}{\bibfnamefont{T.}~\bibnamefont{Strach}},
  \bibinfo{author}{\bibfnamefont{J.}~\bibnamefont{Preston}},
  \bibinfo{author}{\bibfnamefont{T.}~\bibnamefont{Timusk}},
  \bibinfo{author}{\bibfnamefont{D.~A.} \bibnamefont{Bonn}},
  \bibinfo{author}{\bibfnamefont{R.}~\bibnamefont{Liang}}, \bibnamefont{and}
  \bibinfo{author}{\bibfnamefont{W.~N.} \bibnamefont{Hardy}},
  \bibinfo{journal}{Phys. Rev. B} \textbf{\bibinfo{volume}{49}},
  \bibinfo{pages}{12165} (\bibinfo{year}{1994}).

\bibitem[{\citenamefont{Basov et~al.}(1995)\citenamefont{Basov, Liang, Bonn,
  Hardy, Dabrowski, Quijada, Tanner, Rice, Ginsberg, and Timusk}}]{basov95a}
\bibinfo{author}{\bibfnamefont{D.~N.} \bibnamefont{Basov}},
  \bibinfo{author}{\bibfnamefont{R.}~\bibnamefont{Liang}},
  \bibinfo{author}{\bibfnamefont{D.~A.} \bibnamefont{Bonn}},
  \bibinfo{author}{\bibfnamefont{W.~N.} \bibnamefont{Hardy}},
  \bibinfo{author}{\bibfnamefont{B.}~\bibnamefont{Dabrowski}},
  \bibinfo{author}{\bibfnamefont{M.}~\bibnamefont{Quijada}},
  \bibinfo{author}{\bibfnamefont{D.~B.} \bibnamefont{Tanner}},
  \bibinfo{author}{\bibfnamefont{J.~P.} \bibnamefont{Rice}},
  \bibinfo{author}{\bibfnamefont{D.~M.} \bibnamefont{Ginsberg}},
  \bibnamefont{and} \bibinfo{author}{\bibfnamefont{T.}~\bibnamefont{Timusk}},
  \bibinfo{journal}{Phys. Rev. Lett.} \textbf{\bibinfo{volume}{74}},
  \bibinfo{pages}{598} (\bibinfo{year}{1995}).

\bibitem[{\citenamefont{Homes et~al.}(1999)\citenamefont{Homes, Bonn, Liang,
  Hardy, Basov, Timusk, and Clayman}}]{homes99}
\bibinfo{author}{\bibfnamefont{C.~C.} \bibnamefont{Homes}},
  \bibinfo{author}{\bibfnamefont{D.~A.} \bibnamefont{Bonn}},
  \bibinfo{author}{\bibfnamefont{R.}~\bibnamefont{Liang}},
  \bibinfo{author}{\bibfnamefont{W.~N.} \bibnamefont{Hardy}},
  \bibinfo{author}{\bibfnamefont{D.~N.} \bibnamefont{Basov}},
  \bibinfo{author}{\bibfnamefont{T.}~\bibnamefont{Timusk}}, \bibnamefont{and}
  \bibinfo{author}{\bibfnamefont{B.~P.} \bibnamefont{Clayman}},
  \bibinfo{journal}{Phys. Rev. B} \textbf{\bibinfo{volume}{60}},
  \bibinfo{pages}{9782} (\bibinfo{year}{1999}).

\bibitem[{\citenamefont{Liu et~al.}(1990)\citenamefont{Liu, Quijada, Zibold,
  Yoon, Tanner, Cao, Crow, Berger, Margaritondo, Forro et~al.}}]{liu99}
\bibinfo{author}{\bibfnamefont{H.~L.} \bibnamefont{Liu}},
  \bibinfo{author}{\bibfnamefont{M.~A.} \bibnamefont{Quijada}},
  \bibinfo{author}{\bibfnamefont{A.~M.} \bibnamefont{Zibold}},
  \bibinfo{author}{\bibfnamefont{Y.-D.} \bibnamefont{Yoon}},
  \bibinfo{author}{\bibfnamefont{D.~B.} \bibnamefont{Tanner}},
  \bibinfo{author}{\bibfnamefont{G.}~\bibnamefont{Cao}},
  \bibinfo{author}{\bibfnamefont{J.~E.} \bibnamefont{Crow}},
  \bibinfo{author}{\bibfnamefont{H.}~\bibnamefont{Berger}},
  \bibinfo{author}{\bibfnamefont{G.}~\bibnamefont{Margaritondo}},
  \bibinfo{author}{\bibfnamefont{L.}~\bibnamefont{Forro}},
  \bibnamefont{et~al.}, \bibinfo{journal}{J. Phys: Condens. Matter}
  \textbf{\bibinfo{volume}{11}}, \bibinfo{pages}{239} (\bibinfo{year}{1990}).

\bibitem[{\citenamefont{Tu et~al.}(2002)\citenamefont{Tu, Homes, Gu, Basov, and
  Strongin}}]{tu02}
\bibinfo{author}{\bibfnamefont{J.~J.} \bibnamefont{Tu}},
  \bibinfo{author}{\bibfnamefont{C.~C.} \bibnamefont{Homes}},
  \bibinfo{author}{\bibfnamefont{G.~D.} \bibnamefont{Gu}},
  \bibinfo{author}{\bibfnamefont{D.~N.} \bibnamefont{Basov}}, \bibnamefont{and}
  \bibinfo{author}{\bibfnamefont{M.}~\bibnamefont{Strongin}},
  \bibinfo{journal}{Phys. Rev. B} \textbf{\bibinfo{volume}{66}},
  \bibinfo{pages}{144514} (\bibinfo{year}{2002}).

\bibitem[{\citenamefont{Wang et~al.}(1999)\citenamefont{Wang, McConnell, and
  Clayman}}]{wang99}
\bibinfo{author}{\bibfnamefont{N.~L.} \bibnamefont{Wang}},
  \bibinfo{author}{\bibfnamefont{A.~W.} \bibnamefont{McConnell}},
  \bibnamefont{and} \bibinfo{author}{\bibfnamefont{B.~P.}
  \bibnamefont{Clayman}}, \bibinfo{journal}{Phys. Rev. B}
  \textbf{\bibinfo{volume}{60}}, \bibinfo{pages}{14883} (\bibinfo{year}{1999}).

\bibitem[{\citenamefont{Startseva et~al.}(1999)\citenamefont{Startseva, Timusk,
  Puchkov, Basov, Mook, Okuya, Kimura, and Kishio}}]{startseva99}
\bibinfo{author}{\bibfnamefont{T.}~\bibnamefont{Startseva}},
  \bibinfo{author}{\bibfnamefont{T.}~\bibnamefont{Timusk}},
  \bibinfo{author}{\bibfnamefont{A.~V.} \bibnamefont{Puchkov}},
  \bibinfo{author}{\bibfnamefont{D.~N.} \bibnamefont{Basov}},
  \bibinfo{author}{\bibfnamefont{H.~A.} \bibnamefont{Mook}},
  \bibinfo{author}{\bibfnamefont{M.}~\bibnamefont{Okuya}},
  \bibinfo{author}{\bibfnamefont{T.}~\bibnamefont{Kimura}}, \bibnamefont{and}
  \bibinfo{author}{\bibfnamefont{K.}~\bibnamefont{Kishio}},
  \bibinfo{journal}{Phys. Rev. B} \textbf{\bibinfo{volume}{59}},
  \bibinfo{pages}{7184} (\bibinfo{year}{1999}).

\bibitem[{\citenamefont{Puchkov et~al.}(1995)\citenamefont{Puchkov, Timusk,
  Doyle, and Herman}}]{puchkov95}
\bibinfo{author}{\bibfnamefont{A.~V.} \bibnamefont{Puchkov}},
  \bibinfo{author}{\bibfnamefont{T.}~\bibnamefont{Timusk}},
  \bibinfo{author}{\bibfnamefont{S.}~\bibnamefont{Doyle}}, \bibnamefont{and}
  \bibinfo{author}{\bibfnamefont{A.~M.} \bibnamefont{Herman}},
  \bibinfo{journal}{Phys. Rev. B} \textbf{\bibinfo{volume}{51}},
  \bibinfo{pages}{3312} (\bibinfo{year}{1995}).

\bibitem[{\citenamefont{Singley et~al.}(2001)\citenamefont{Singley, Basov,
  Kurahashi, Uefuji, and Yamada}}]{singley01}
\bibinfo{author}{\bibfnamefont{E.~J.} \bibnamefont{Singley}},
  \bibinfo{author}{\bibfnamefont{D.~N.} \bibnamefont{Basov}},
  \bibinfo{author}{\bibfnamefont{K.}~\bibnamefont{Kurahashi}},
  \bibinfo{author}{\bibfnamefont{T.}~\bibnamefont{Uefuji}}, \bibnamefont{and}
  \bibinfo{author}{\bibfnamefont{K.}~\bibnamefont{Yamada}},
  \bibinfo{journal}{Phys. Rev. B} \textbf{\bibinfo{volume}{64}},
  \bibinfo{pages}{224503} (\bibinfo{year}{2001}).

\bibitem[{\citenamefont{Zimmers et~al.}(2004)\citenamefont{Zimmers, Lobo,
  Bontemps, Homes, Barr, Dagan, and Greene}}]{zimmers04}
\bibinfo{author}{\bibfnamefont{A.}~\bibnamefont{Zimmers}},
  \bibinfo{author}{\bibfnamefont{R.}~\bibnamefont{Lobo}},
  \bibinfo{author}{\bibfnamefont{N.}~\bibnamefont{Bontemps}},
  \bibinfo{author}{\bibfnamefont{C.}~\bibnamefont{Homes}},
  \bibinfo{author}{\bibfnamefont{M.}~\bibnamefont{Barr}},
  \bibinfo{author}{\bibfnamefont{Y.}~\bibnamefont{Dagan}}, \bibnamefont{and}
  \bibinfo{author}{\bibfnamefont{R.}~\bibnamefont{Greene}},
  \bibinfo{journal}{Phys. Rev. B} \textbf{\bibinfo{volume}{70}},
  \bibinfo{pages}{132502} (\bibinfo{year}{2004}).

\bibitem[{\citenamefont{Puchkov
  et~al.}(1996{\natexlab{a}})\citenamefont{Puchkov, Timusk, Karlow, Cooper,
  Han, and Payne}}]{puchkov96b}
\bibinfo{author}{\bibfnamefont{A.~V.} \bibnamefont{Puchkov}},
  \bibinfo{author}{\bibfnamefont{T.}~\bibnamefont{Timusk}},
  \bibinfo{author}{\bibfnamefont{M.~A.} \bibnamefont{Karlow}},
  \bibinfo{author}{\bibfnamefont{S.~L.} \bibnamefont{Cooper}},
  \bibinfo{author}{\bibfnamefont{P.~D.} \bibnamefont{Han}}, \bibnamefont{and}
  \bibinfo{author}{\bibfnamefont{D.~A.} \bibnamefont{Payne}},
  \bibinfo{journal}{Phys. Rev. B} \textbf{\bibinfo{volume}{54}},
  \bibinfo{pages}{6686} (\bibinfo{year}{1996}{\natexlab{a}}).

\bibitem[{\citenamefont{Ferrell and Schmidt}(1967)}]{ferrell67}
\bibinfo{author}{\bibfnamefont{R.~A.} \bibnamefont{Ferrell}} \bibnamefont{and}
  \bibinfo{author}{\bibfnamefont{H.}~\bibnamefont{Schmidt}},
  \bibinfo{journal}{Phys. Lett.} \textbf{\bibinfo{volume}{25A}},
  \bibinfo{pages}{544} (\bibinfo{year}{1967}).

\bibitem[{\citenamefont{Klein et~al.}(1994)\citenamefont{Klein, Nicol, Holczer,
  and Gr{\"u}ner}}]{klein94}
\bibinfo{author}{\bibfnamefont{O.}~\bibnamefont{Klein}},
  \bibinfo{author}{\bibfnamefont{E.~J.} \bibnamefont{Nicol}},
  \bibinfo{author}{\bibfnamefont{K.}~\bibnamefont{Holczer}}, \bibnamefont{and}
  \bibinfo{author}{\bibfnamefont{G.}~\bibnamefont{Gr{\"u}ner}},
  \bibinfo{journal}{Phys. Rev. B} \textbf{\bibinfo{volume}{50}},
  \bibinfo{pages}{6307} (\bibinfo{year}{1994}).

\bibitem[{\citenamefont{Pronin et~al.}(1998)\citenamefont{Pronin, Dressel,
  Pimonev, Loidl, Roshchin, and Greene}}]{pronin98}
\bibinfo{author}{\bibfnamefont{A.~V.} \bibnamefont{Pronin}},
  \bibinfo{author}{\bibfnamefont{M.}~\bibnamefont{Dressel}},
  \bibinfo{author}{\bibfnamefont{A.}~\bibnamefont{Pimonev}},
  \bibinfo{author}{\bibfnamefont{A.}~\bibnamefont{Loidl}},
  \bibinfo{author}{\bibfnamefont{I.~V.} \bibnamefont{Roshchin}},
  \bibnamefont{and} \bibinfo{author}{\bibfnamefont{L.~H.}
  \bibnamefont{Greene}}, \bibinfo{journal}{Phys. Rev. B}
  \textbf{\bibinfo{volume}{57}}, \bibinfo{pages}{14416} (\bibinfo{year}{1998}).

\bibitem[{\citenamefont{Shen et~al.}(1993)\citenamefont{Shen, Dessau, Wells,
  King, Spicer, Arko, Marshall, Lombardo, Kapitulnik, Dickinson
  et~al.}}]{shen93}
\bibinfo{author}{\bibfnamefont{Z.}~\bibnamefont{Shen}},
  \bibinfo{author}{\bibfnamefont{D.~S.} \bibnamefont{Dessau}},
  \bibinfo{author}{\bibfnamefont{B.~O.} \bibnamefont{Wells}},
  \bibinfo{author}{\bibfnamefont{D.~M.} \bibnamefont{King}},
  \bibinfo{author}{\bibfnamefont{W.~E.} \bibnamefont{Spicer}},
  \bibinfo{author}{\bibfnamefont{A.~J.} \bibnamefont{Arko}},
  \bibinfo{author}{\bibfnamefont{D.}~\bibnamefont{Marshall}},
  \bibinfo{author}{\bibfnamefont{L.~W.} \bibnamefont{Lombardo}},
  \bibinfo{author}{\bibfnamefont{A.}~\bibnamefont{Kapitulnik}},
  \bibinfo{author}{\bibfnamefont{P.}~\bibnamefont{Dickinson}},
  \bibnamefont{et~al.}, \bibinfo{journal}{Phys. Rev. Lett.}
  \textbf{\bibinfo{volume}{70}}, \bibinfo{pages}{1553} (\bibinfo{year}{1993}).

\bibitem[{\citenamefont{Hardy et~al.}(1993)\citenamefont{Hardy, Bonn, Morgan,
  Liang, and Zhang}}]{hardy93}
\bibinfo{author}{\bibfnamefont{W.~N.} \bibnamefont{Hardy}},
  \bibinfo{author}{\bibfnamefont{D.~A.} \bibnamefont{Bonn}},
  \bibinfo{author}{\bibfnamefont{D.~C.} \bibnamefont{Morgan}},
  \bibinfo{author}{\bibfnamefont{R.}~\bibnamefont{Liang}}, \bibnamefont{and}
  \bibinfo{author}{\bibfnamefont{K.}~\bibnamefont{Zhang}},
  \bibinfo{journal}{Phys. Rev. Lett.} \textbf{\bibinfo{volume}{70}},
  \bibinfo{pages}{3999} (\bibinfo{year}{1993}).

\bibitem[{\citenamefont{Iye}(1992)}]{iye92}
\bibinfo{author}{\bibfnamefont{Y.}~\bibnamefont{Iye}}, in
  \emph{\bibinfo{booktitle}{Physical Properties of High-Temperature
  Superconductors III}}, edited by \bibinfo{editor}{\bibfnamefont{D.~M.}
  \bibnamefont{Ginsberg}} (\bibinfo{publisher}{World Scientific},
  \bibinfo{address}{Singapore}, \bibinfo{year}{1992}), pp.
  \bibinfo{pages}{285--361}.

\bibitem[{\citenamefont{Bonn et~al.}(1993)\citenamefont{Bonn, Liang, Riseman,
  Baar, Morgan, Zhang, Dosanjh, Duty, MacFarlane, Morris et~al.}}]{bonn93}
\bibinfo{author}{\bibfnamefont{D.~A.} \bibnamefont{Bonn}},
  \bibinfo{author}{\bibfnamefont{R.}~\bibnamefont{Liang}},
  \bibinfo{author}{\bibfnamefont{T.~M.} \bibnamefont{Riseman}},
  \bibinfo{author}{\bibfnamefont{D.~J.} \bibnamefont{Baar}},
  \bibinfo{author}{\bibfnamefont{D.~C.} \bibnamefont{Morgan}},
  \bibinfo{author}{\bibfnamefont{K.}~\bibnamefont{Zhang}},
  \bibinfo{author}{\bibfnamefont{P.}~\bibnamefont{Dosanjh}},
  \bibinfo{author}{\bibfnamefont{T.~L.} \bibnamefont{Duty}},
  \bibinfo{author}{\bibfnamefont{A.}~\bibnamefont{MacFarlane}},
  \bibinfo{author}{\bibfnamefont{G.~D.} \bibnamefont{Morris}},
  \bibnamefont{et~al.}, \bibinfo{journal}{Phys. Rev. B}
  \textbf{\bibinfo{volume}{47}}, \bibinfo{pages}{11314} (\bibinfo{year}{1993}).

\bibitem[{\citenamefont{Puchkov
  et~al.}(1996{\natexlab{b}})\citenamefont{Puchkov, Basov, and
  Timusk}}]{puchkov96a}
\bibinfo{author}{\bibfnamefont{A.~V.} \bibnamefont{Puchkov}},
  \bibinfo{author}{\bibfnamefont{D.~N.} \bibnamefont{Basov}}, \bibnamefont{and}
  \bibinfo{author}{\bibfnamefont{T.}~\bibnamefont{Timusk}},
  \bibinfo{journal}{J. Phys: Condens. Matter} \textbf{\bibinfo{volume}{8}},
  \bibinfo{pages}{10049} (\bibinfo{year}{1996}{\natexlab{b}}).

\bibitem[{\citenamefont{Sutherland et~al.}(2003)\citenamefont{Sutherland,
  Hawthorn, Hill, Ronning, Wakimoto, Zhang, Proust, Boaknin, Lupien, Taillefer
  et~al.}}]{sutherland03}
\bibinfo{author}{\bibfnamefont{M.}~\bibnamefont{Sutherland}},
  \bibinfo{author}{\bibfnamefont{D.~G.} \bibnamefont{Hawthorn}},
  \bibinfo{author}{\bibfnamefont{R.~W.} \bibnamefont{Hill}},
  \bibinfo{author}{\bibfnamefont{R.}~\bibnamefont{Ronning}},
  \bibinfo{author}{\bibfnamefont{S.}~\bibnamefont{Wakimoto}},
  \bibinfo{author}{\bibfnamefont{H.}~\bibnamefont{Zhang}},
  \bibinfo{author}{\bibfnamefont{C.}~\bibnamefont{Proust}},
  \bibinfo{author}{\bibfnamefont{E.}~\bibnamefont{Boaknin}},
  \bibinfo{author}{\bibfnamefont{C.}~\bibnamefont{Lupien}},
  \bibinfo{author}{\bibfnamefont{L.}~\bibnamefont{Taillefer}},
  \bibnamefont{et~al.}, \bibinfo{journal}{Phys. Rev. B}
  \textbf{\bibinfo{volume}{67}}, \bibinfo{pages}{174520}
  (\bibinfo{year}{2003}).

\bibitem[{\citenamefont{Turner et~al.}(2003)\citenamefont{Turner, Harris,
  Kamal, Hayden, Broun, Morgan, Hosseini, Dosanjh, Mullins, Preston
  et~al.}}]{turner03}
\bibinfo{author}{\bibfnamefont{P.~J.} \bibnamefont{Turner}},
  \bibinfo{author}{\bibfnamefont{R.}~\bibnamefont{Harris}},
  \bibinfo{author}{\bibfnamefont{S.}~\bibnamefont{Kamal}},
  \bibinfo{author}{\bibfnamefont{M.~E.} \bibnamefont{Hayden}},
  \bibinfo{author}{\bibfnamefont{D.~M.} \bibnamefont{Broun}},
  \bibinfo{author}{\bibfnamefont{D.~C.} \bibnamefont{Morgan}},
  \bibinfo{author}{\bibfnamefont{A.}~\bibnamefont{Hosseini}},
  \bibinfo{author}{\bibfnamefont{P.}~\bibnamefont{Dosanjh}},
  \bibinfo{author}{\bibfnamefont{G.~K.} \bibnamefont{Mullins}},
  \bibinfo{author}{\bibfnamefont{J.~S.} \bibnamefont{Preston}},
  \bibnamefont{et~al.}, \bibinfo{journal}{Phys. Rev. Lett.}
  \textbf{\bibinfo{volume}{90}}, \bibinfo{pages}{237005}
  (\bibinfo{year}{2003}).

\bibitem[{\citenamefont{Hill et~al.}(2004)\citenamefont{Hill, Lupien,
  Sutherland, Boaknin, Hawthorn, Proust, Ronning, Taillefer, Liang, Bonn
  et~al.}}]{hill04}
\bibinfo{author}{\bibfnamefont{R.~W.} \bibnamefont{Hill}},
  \bibinfo{author}{\bibfnamefont{C.}~\bibnamefont{Lupien}},
  \bibinfo{author}{\bibfnamefont{M.}~\bibnamefont{Sutherland}},
  \bibinfo{author}{\bibfnamefont{E.}~\bibnamefont{Boaknin}},
  \bibinfo{author}{\bibfnamefont{D.~G.} \bibnamefont{Hawthorn}},
  \bibinfo{author}{\bibfnamefont{C.}~\bibnamefont{Proust}},
  \bibinfo{author}{\bibfnamefont{F.}~\bibnamefont{Ronning}},
  \bibinfo{author}{\bibfnamefont{L.}~\bibnamefont{Taillefer}},
  \bibinfo{author}{\bibfnamefont{R.}~\bibnamefont{Liang}},
  \bibinfo{author}{\bibfnamefont{D.~A.} \bibnamefont{Bonn}},
  \bibnamefont{et~al.}, \bibinfo{journal}{Phys. Rev. Lett.}
  \textbf{\bibinfo{volume}{92}}, \bibinfo{pages}{027001}
  (\bibinfo{year}{2004}).

\bibitem[{\citenamefont{Basov et~al.}(2002)\citenamefont{Basov, Singley, and
  Dordevic}}]{basov02}
\bibinfo{author}{\bibfnamefont{D.~N.} \bibnamefont{Basov}},
  \bibinfo{author}{\bibfnamefont{E.~J.} \bibnamefont{Singley}},
  \bibnamefont{and} \bibinfo{author}{\bibfnamefont{S.~V.}
  \bibnamefont{Dordevic}}, \bibinfo{journal}{Phys. Rev. B}
  \textbf{\bibinfo{volume}{65}}, \bibinfo{pages}{054516}
  (\bibinfo{year}{2002}).

\bibitem[{\citenamefont{Boebinger et~al.}(1996)\citenamefont{Boebinger, Ando,
  Passner, Kimura, Okura, Shimoyama, Kishio, Tamasaku, Ichikawa, and
  Uchida}}]{boebinger96}
\bibinfo{author}{\bibfnamefont{G.~S.} \bibnamefont{Boebinger}},
  \bibinfo{author}{\bibfnamefont{Y.}~\bibnamefont{Ando}},
  \bibinfo{author}{\bibfnamefont{A.}~\bibnamefont{Passner}},
  \bibinfo{author}{\bibfnamefont{T.}~\bibnamefont{Kimura}},
  \bibinfo{author}{\bibfnamefont{M.}~\bibnamefont{Okura}},
  \bibinfo{author}{\bibfnamefont{J.}~\bibnamefont{Shimoyama}},
  \bibinfo{author}{\bibfnamefont{K.}~\bibnamefont{Kishio}},
  \bibinfo{author}{\bibfnamefont{K.}~\bibnamefont{Tamasaku}},
  \bibinfo{author}{\bibfnamefont{N.}~\bibnamefont{Ichikawa}}, \bibnamefont{and}
  \bibinfo{author}{\bibfnamefont{S.}~\bibnamefont{Uchida}},
  \bibinfo{journal}{Phys. Rev. Lett.} \textbf{\bibinfo{volume}{77}},
  \bibinfo{pages}{5417} (\bibinfo{year}{1996}).

\bibitem[{\citenamefont{Ando et~al.}(1996)\citenamefont{Ando, Boebinger,
  Passner, Wang, Geibel, and Steglich}}]{ando96}
\bibinfo{author}{\bibfnamefont{Y.}~\bibnamefont{Ando}},
  \bibinfo{author}{\bibfnamefont{G.~S.} \bibnamefont{Boebinger}},
  \bibinfo{author}{\bibfnamefont{A.}~\bibnamefont{Passner}},
  \bibinfo{author}{\bibfnamefont{N.~L.} \bibnamefont{Wang}},
  \bibinfo{author}{\bibfnamefont{C.}~\bibnamefont{Geibel}}, \bibnamefont{and}
  \bibinfo{author}{\bibfnamefont{F.}~\bibnamefont{Steglich}},
  \bibinfo{journal}{Phys. Rev. Lett.} \textbf{\bibinfo{volume}{77}},
  \bibinfo{pages}{2065} (\bibinfo{year}{1996}).

\bibitem[{\citenamefont{Ando et~al.}(1997{\natexlab{a}})\citenamefont{Ando,
  Boebinger, Passner, Wang, Geibel, and Steglich}}]{ando97a}
\bibinfo{author}{\bibfnamefont{Y.}~\bibnamefont{Ando}},
  \bibinfo{author}{\bibfnamefont{G.~S.} \bibnamefont{Boebinger}},
  \bibinfo{author}{\bibfnamefont{A.}~\bibnamefont{Passner}},
  \bibinfo{author}{\bibfnamefont{N.~L.} \bibnamefont{Wang}},
  \bibinfo{author}{\bibfnamefont{C.}~\bibnamefont{Geibel}}, \bibnamefont{and}
  \bibinfo{author}{\bibfnamefont{F.}~\bibnamefont{Steglich}},
  \bibinfo{journal}{Phys. Rev. Lett.} \textbf{\bibinfo{volume}{79}},
  \bibinfo{pages}{2595} (\bibinfo{year}{1997}{\natexlab{a}}).

\bibitem[{\citenamefont{Ando et~al.}(1997{\natexlab{b}})\citenamefont{Ando,
  Boebinger, Passner, Wang, Geibel, Steglich, Trofimov, and
  Balakirev}}]{ando97b}
\bibinfo{author}{\bibfnamefont{Y.}~\bibnamefont{Ando}},
  \bibinfo{author}{\bibfnamefont{G.~S.} \bibnamefont{Boebinger}},
  \bibinfo{author}{\bibfnamefont{A.}~\bibnamefont{Passner}},
  \bibinfo{author}{\bibfnamefont{N.~L.} \bibnamefont{Wang}},
  \bibinfo{author}{\bibfnamefont{C.}~\bibnamefont{Geibel}},
  \bibinfo{author}{\bibfnamefont{F.}~\bibnamefont{Steglich}},
  \bibinfo{author}{\bibfnamefont{I.~E.} \bibnamefont{Trofimov}},
  \bibnamefont{and} \bibinfo{author}{\bibfnamefont{F.~F.}
  \bibnamefont{Balakirev}}, \bibinfo{journal}{Phys. Rev. B}
  \textbf{\bibinfo{volume}{56}}, \bibinfo{pages}{R8530}
  (\bibinfo{year}{1997}{\natexlab{b}}).

\bibitem[{\citenamefont{Ono et~al.}(2000)\citenamefont{Ono, Ando, Murayama,
  Balakirev, Betts, and Boebinger}}]{ono00}
\bibinfo{author}{\bibfnamefont{S.}~\bibnamefont{Ono}},
  \bibinfo{author}{\bibfnamefont{Y.}~\bibnamefont{Ando}},
  \bibinfo{author}{\bibfnamefont{T.}~\bibnamefont{Murayama}},
  \bibinfo{author}{\bibfnamefont{F.~F.} \bibnamefont{Balakirev}},
  \bibinfo{author}{\bibfnamefont{J.~B.} \bibnamefont{Betts}}, \bibnamefont{and}
  \bibinfo{author}{\bibfnamefont{G.~S.} \bibnamefont{Boebinger}},
  \bibinfo{journal}{Phys. Rev. Lett.} \textbf{\bibinfo{volume}{85}},
  \bibinfo{pages}{638} (\bibinfo{year}{2000}).

\bibitem[{\citenamefont{Zimmerman et~al.}(1991)\citenamefont{Zimmerman, Brandt,
  Bauer, Seider, and Genzel}}]{zimmerman91}
\bibinfo{author}{\bibfnamefont{W.}~\bibnamefont{Zimmerman}},
  \bibinfo{author}{\bibfnamefont{E.~H.} \bibnamefont{Brandt}},
  \bibinfo{author}{\bibfnamefont{M.}~\bibnamefont{Bauer}},
  \bibinfo{author}{\bibfnamefont{E.}~\bibnamefont{Seider}}, \bibnamefont{and}
  \bibinfo{author}{\bibfnamefont{L.}~\bibnamefont{Genzel}},
  \bibinfo{journal}{Physica C} \textbf{\bibinfo{volume}{183}},
  \bibinfo{pages}{99} (\bibinfo{year}{1991}).

\bibitem[{\citenamefont{Orenstein et~al.}(1990)\citenamefont{Orenstein, Thomas,
  Millis, Cooper, Rapkine, Timusk, Schneemeyer, and Waszczak}}]{orenstein90}
\bibinfo{author}{\bibfnamefont{J.}~\bibnamefont{Orenstein}},
  \bibinfo{author}{\bibfnamefont{G.~A.} \bibnamefont{Thomas}},
  \bibinfo{author}{\bibfnamefont{A.~J.} \bibnamefont{Millis}},
  \bibinfo{author}{\bibfnamefont{S.~L.} \bibnamefont{Cooper}},
  \bibinfo{author}{\bibfnamefont{D.~H.} \bibnamefont{Rapkine}},
  \bibinfo{author}{\bibfnamefont{T.}~\bibnamefont{Timusk}},
  \bibinfo{author}{\bibfnamefont{L.~F.} \bibnamefont{Schneemeyer}},
  \bibnamefont{and} \bibinfo{author}{\bibfnamefont{J.~V.}
  \bibnamefont{Waszczak}}, \bibinfo{journal}{Phys. Rev. B}
  \textbf{\bibinfo{volume}{42}}, \bibinfo{pages}{6342} (\bibinfo{year}{1990}).

\bibitem[{\citenamefont{Zaanen}(2004)}]{zaanen04}
\bibinfo{author}{\bibfnamefont{J.}~\bibnamefont{Zaanen}},
  \bibinfo{journal}{Nature} \textbf{\bibinfo{volume}{430}},
  \bibinfo{pages}{512} (\bibinfo{year}{2004}).

\bibitem[{\citenamefont{Lee et~al.}(2004)\citenamefont{Lee, Segawa, Ando, and
  Basov}}]{lee04}
\bibinfo{author}{\bibfnamefont{Y.-S.} \bibnamefont{Lee}},
  \bibinfo{author}{\bibfnamefont{K.}~\bibnamefont{Segawa}},
  \bibinfo{author}{\bibfnamefont{Y.}~\bibnamefont{Ando}}, \bibnamefont{and}
  \bibinfo{author}{\bibfnamefont{D.~N.} \bibnamefont{Basov}},
  \bibinfo{journal}{Phys. Rev. B} \textbf{\bibinfo{volume}{70}},
  \bibinfo{pages}{014518} (\bibinfo{year}{2004}).

\bibitem[{\citenamefont{Hardy et~al.}(1994)\citenamefont{Hardy, Kamal, Bonn,
  Zhang, Liang, Klein, Morgan, and Baar}}]{hardy94}
\bibinfo{author}{\bibfnamefont{W.~N.} \bibnamefont{Hardy}},
  \bibinfo{author}{\bibfnamefont{S.}~\bibnamefont{Kamal}},
  \bibinfo{author}{\bibfnamefont{D.~A.} \bibnamefont{Bonn}},
  \bibinfo{author}{\bibfnamefont{K.}~\bibnamefont{Zhang}},
  \bibinfo{author}{\bibfnamefont{R.}~\bibnamefont{Liang}},
  \bibinfo{author}{\bibfnamefont{E.}~\bibnamefont{Klein}},
  \bibinfo{author}{\bibfnamefont{D.~C.} \bibnamefont{Morgan}},
  \bibnamefont{and} \bibinfo{author}{\bibfnamefont{D.}~\bibnamefont{Baar}},
  \bibinfo{journal}{Physica B} \textbf{\bibinfo{volume}{197}},
  \bibinfo{pages}{609} (\bibinfo{year}{1994}).

\bibitem[{\citenamefont{Hosseini et~al.}(1999)\citenamefont{Hosseini, Harris,
  Kamal, Dosanjh, Preston, Liang, Hardy, and Bonn}}]{hosseini99}
\bibinfo{author}{\bibfnamefont{A.}~\bibnamefont{Hosseini}},
  \bibinfo{author}{\bibfnamefont{R.}~\bibnamefont{Harris}},
  \bibinfo{author}{\bibfnamefont{S.}~\bibnamefont{Kamal}},
  \bibinfo{author}{\bibfnamefont{P.}~\bibnamefont{Dosanjh}},
  \bibinfo{author}{\bibfnamefont{J.}~\bibnamefont{Preston}},
  \bibinfo{author}{\bibfnamefont{R.}~\bibnamefont{Liang}},
  \bibinfo{author}{\bibfnamefont{W.~N.} \bibnamefont{Hardy}}, \bibnamefont{and}
  \bibinfo{author}{\bibfnamefont{D.~A.} \bibnamefont{Bonn}},
  \bibinfo{journal}{Phys. Rev. B} \textbf{\bibinfo{volume}{60}},
  \bibinfo{pages}{1349} (\bibinfo{year}{1999}).

\bibitem[{\citenamefont{Varmazis and Strongin}(1974)}]{varmazis74}
\bibinfo{author}{\bibfnamefont{C.}~\bibnamefont{Varmazis}} \bibnamefont{and}
  \bibinfo{author}{\bibfnamefont{M.}~\bibnamefont{Strongin}},
  \bibinfo{journal}{Phys. Rev. B} \textbf{\bibinfo{volume}{10}},
  \bibinfo{pages}{1885} (\bibinfo{year}{1974}).

\bibitem[{\citenamefont{Palik}(1998)}]{palik}
\bibinfo{editor}{\bibfnamefont{E.~D.} \bibnamefont{Palik}}, ed.,
  \emph{\bibinfo{title}{Handbook of Optical Constants of Solids II}}
  (\bibinfo{publisher}{Academic Press}, \bibinfo{address}{San Diego},
  \bibinfo{year}{1998}).

\bibitem[{\citenamefont{Smith and Ambegaokar}(1992)}]{smith92}
\bibinfo{author}{\bibfnamefont{R.~A.} \bibnamefont{Smith}} \bibnamefont{and}
  \bibinfo{author}{\bibfnamefont{V.}~\bibnamefont{Ambegaokar}},
  \bibinfo{journal}{Phys. Rev. B} \textbf{\bibinfo{volume}{45}},
  \bibinfo{pages}{2463} (\bibinfo{year}{1992}), \bibinfo{note}{in the absence
  of weak localization, in the dirty limit $\Gamma \gg \Delta$, it was shown
  that $\rho_s = 4\pi^2\sigma_{dc}\Delta \tanh [\Delta /2T]$, where
  $\sigma_{dc}$ is the dc conductivity in the normal state and $\Delta$ is the
  superconducting energy gap. Assuming that the gap scales with $T_c$ and that
  $T\ll T_c$, then $\rho_s \propto \sigma_{dc} T_c$. This result is similar to
  the one obtained in the text using arguments based on the spectral weight.}

\bibitem[{\citenamefont{Panagopoulos and Xiang}(1998)}]{panagopoulos98}
\bibinfo{author}{\bibfnamefont{C.}~\bibnamefont{Panagopoulos}}
  \bibnamefont{and} \bibinfo{author}{\bibfnamefont{T.}~\bibnamefont{Xiang}},
  \bibinfo{journal}{Phys. Rev. Lett.} \textbf{\bibinfo{volume}{81}},
  \bibinfo{pages}{2336} (\bibinfo{year}{1998}).

\bibitem[{\citenamefont{Damascelli et~al.}(2003)\citenamefont{Damascelli,
  Hussain, and Shen}}]{damascelli03}
\bibinfo{author}{\bibfnamefont{A.}~\bibnamefont{Damascelli}},
  \bibinfo{author}{\bibfnamefont{Z.}~\bibnamefont{Hussain}}, \bibnamefont{and}
  \bibinfo{author}{\bibfnamefont{Z.-X.} \bibnamefont{Shen}},
  \bibinfo{journal}{Rev. Mod. Phys.} \textbf{\bibinfo{volume}{75}},
  \bibinfo{pages}{473} (\bibinfo{year}{2003}).

\bibitem[{\citenamefont{Yeh et~al.}(2001)\citenamefont{Yeh, Chen, Hammerl,
  Mannhart, Schmehl, Schneider, Schulz, Tajima, Yoshida, Garrigus
  et~al.}}]{yeh01}
\bibinfo{author}{\bibfnamefont{N.-C.} \bibnamefont{Yeh}},
  \bibinfo{author}{\bibfnamefont{C.-T.} \bibnamefont{Chen}},
  \bibinfo{author}{\bibfnamefont{G.}~\bibnamefont{Hammerl}},
  \bibinfo{author}{\bibfnamefont{J.}~\bibnamefont{Mannhart}},
  \bibinfo{author}{\bibfnamefont{A.}~\bibnamefont{Schmehl}},
  \bibinfo{author}{\bibfnamefont{C.~W.} \bibnamefont{Schneider}},
  \bibinfo{author}{\bibfnamefont{R.~R.} \bibnamefont{Schulz}},
  \bibinfo{author}{\bibfnamefont{S.}~\bibnamefont{Tajima}},
  \bibinfo{author}{\bibfnamefont{K.}~\bibnamefont{Yoshida}},
  \bibinfo{author}{\bibfnamefont{D.}~\bibnamefont{Garrigus}},
  \bibnamefont{et~al.}, \bibinfo{journal}{Phys. Rev. Lett.}
  \textbf{\bibinfo{volume}{87}}, \bibinfo{pages}{087003}
  (\bibinfo{year}{2001}).

\bibitem[{\citenamefont{Santander-Syro
  et~al.}(2002)\citenamefont{Santander-Syro, Lobo, Bontemps, Konstantinovic,
  Li, and Raffy}}]{syro02}
\bibinfo{author}{\bibfnamefont{A.~F.} \bibnamefont{Santander-Syro}},
  \bibinfo{author}{\bibfnamefont{R.~P. S.~M.} \bibnamefont{Lobo}},
  \bibinfo{author}{\bibfnamefont{N.}~\bibnamefont{Bontemps}},
  \bibinfo{author}{\bibfnamefont{Z.}~\bibnamefont{Konstantinovic}},
  \bibinfo{author}{\bibfnamefont{Z.}~\bibnamefont{Li}}, \bibnamefont{and}
  \bibinfo{author}{\bibfnamefont{H.}~\bibnamefont{Raffy}},
  \bibinfo{journal}{Phys. Rev. Lett} \textbf{\bibinfo{volume}{88}},
  \bibinfo{pages}{097005} (\bibinfo{year}{2002}).

\bibitem[{\citenamefont{Homes et~al.}(2004{\natexlab{b}})\citenamefont{Homes,
  Dordevic, Bonn, Liang, and Hardy}}]{homes04a}
\bibinfo{author}{\bibfnamefont{C.~C.} \bibnamefont{Homes}},
  \bibinfo{author}{\bibfnamefont{S.~V.} \bibnamefont{Dordevic}},
  \bibinfo{author}{\bibfnamefont{D.~A.} \bibnamefont{Bonn}},
  \bibinfo{author}{\bibfnamefont{R.}~\bibnamefont{Liang}}, \bibnamefont{and}
  \bibinfo{author}{\bibfnamefont{W.~N.} \bibnamefont{Hardy}},
  \bibinfo{journal}{Phys. Rev. B} \textbf{\bibinfo{volume}{69}},
  \bibinfo{pages}{024514} (\bibinfo{year}{2004}{\natexlab{b}}).

\bibitem[{\citenamefont{Homes et~al.}()\citenamefont{Homes, Dordevic, Bonn,
  Liang, Hardy, and Timusk}}]{homes03}
\bibinfo{author}{\bibfnamefont{C.~C.} \bibnamefont{Homes}},
  \bibinfo{author}{\bibfnamefont{S.~V.} \bibnamefont{Dordevic}},
  \bibinfo{author}{\bibfnamefont{D.~A.} \bibnamefont{Bonn}},
  \bibinfo{author}{\bibfnamefont{R.}~\bibnamefont{Liang}},
  \bibinfo{author}{\bibfnamefont{W.~N.} \bibnamefont{Hardy}}, \bibnamefont{and}
  \bibinfo{author}{\bibfnamefont{T.}~\bibnamefont{Timusk}},
  \eprint{cond-mat/0312211}.

\bibitem[{\citenamefont{Cooper and Gray}(1994)}]{cooper94}
\bibinfo{author}{\bibfnamefont{S.~L.} \bibnamefont{Cooper}} \bibnamefont{and}
  \bibinfo{author}{\bibfnamefont{K.~E.} \bibnamefont{Gray}}, in
  \emph{\bibinfo{booktitle}{Physical Properties of High-Temperature
  Superconductors IV}}, edited by \bibinfo{editor}{\bibfnamefont{D.~M.}
  \bibnamefont{Ginsberg}} (\bibinfo{publisher}{World Scientific},
  \bibinfo{address}{Singapore}, \bibinfo{year}{1994}), pp.
  \bibinfo{pages}{61--188}.

\bibitem[{\citenamefont{Jung et~al.}(1998)\citenamefont{Jung, Yan, Darhmaoui,
  and Kwok}}]{jung98}
\bibinfo{author}{\bibfnamefont{J.}~\bibnamefont{Jung}},
  \bibinfo{author}{\bibfnamefont{H.}~\bibnamefont{Yan}},
  \bibinfo{author}{\bibfnamefont{H.}~\bibnamefont{Darhmaoui}},
  \bibnamefont{and} \bibinfo{author}{\bibfnamefont{W.-K.} \bibnamefont{Kwok}},
  in \emph{\bibinfo{booktitle}{Superconducting and Related Oxides: Physics and
  Nanoengineering III}}, edited by
  \bibinfo{editor}{\bibfnamefont{D.}~\bibnamefont{Pavuna}} \bibnamefont{and}
  \bibinfo{editor}{\bibfnamefont{I.}~\bibnamefont{Bozovic}}
  (\bibinfo{publisher}{SPIE}, \bibinfo{address}{Washington},
  \bibinfo{year}{1998}), vol. \bibinfo{volume}{3481}, pp.
  \bibinfo{pages}{172--181}.

\bibitem[{\citenamefont{Tu et~al.}()\citenamefont{Tu, Strongin, and
  Imry}}]{tu04}
\bibinfo{author}{\bibfnamefont{J.~J.} \bibnamefont{Tu}},
  \bibinfo{author}{\bibfnamefont{M.}~\bibnamefont{Strongin}}, \bibnamefont{and}
  \bibinfo{author}{\bibfnamefont{Y.}~\bibnamefont{Imry}},
  \eprint{cond-mat/0405625}.

\end{thebibliography}

\end{document}